\documentclass[preprint,review,12pt]{elsarticle}
\pdfoutput=1

\usepackage{amssymb}
\usepackage{amsthm}
\usepackage{amsmath}
\usepackage{upgreek}
\usepackage{listings}
\usepackage{multirow}
\usepackage{multicol}
\usepackage{color}
\usepackage{xcolor}
\usepackage{stmaryrd}
\usepackage{booktabs}
\usepackage{calc}
\usepackage{soul}
\usepackage[section]{placeins}
\usepackage[titletoc]{appendix}
\usepackage[binary-units=true]{siunitx}
\usepackage[ruled,vlined]{algorithm2e}

\usepackage{graphicx}
\usepackage{graphics}
\usepackage{float}
\usepackage{subcaption}
\usepackage{wrapfig}
\usepackage{varwidth}
\usepackage{tikz}
\usepackage{pgfplots}
\usepackage{tikz-layers}
\usetikzlibrary{decorations.text}
\usetikzlibrary{arrows,matrix,positioning,fit}
\usetikzlibrary{shapes,positioning}
\usetikzlibrary{backgrounds}
\pgfplotsset{compat=1.14}
\usepgfplotslibrary{colorbrewer}
\usepgfplotslibrary{patchplots}
\usepgfplotslibrary[colorbrewer]
\usetikzlibrary{pgfplots.colorbrewer}
\usetikzlibrary[pgfplots.colorbrewer]
\usepgfplotslibrary{units}
\usetikzlibrary{spy}
\usepgfplotslibrary{patchplots}
\usepgfplotslibrary[colorbrewer]
\usepgfplotslibrary{fillbetween}
\usetikzlibrary{decorations.pathmorphing,calc}
\usetikzlibrary{intersections}
\usepackage{pgfplotstable}
\usepackage{arrayjobx}
\graphicspath{ {./Figs/} }

\newcommand{\mathhl}[2]{\colorbox{#1}{$\phantom{[}\displaystyle #2$}}

\newlength\myheight
\newlength\mydepth
\settototalheight\myheight{Xygp}
\newcommand*\inlinegraphics[1]{%
  \settototalheight\myheight{Xygp}%
  \settodepth\mydepth{Xygp}%
  \raisebox{-\mydepth}{\includegraphics[height=\myheight]{#1}}%
}
\newcommand\orcid[1]{\href{https://orcid.org/#1}{\inlinegraphics{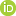}}}

\makeatletter
\def\BState{\State\hskip-\ALG@thistlm}
\makeatother

\newcommand{\etal}{{et~al.}}
\newcommand\floor[1]{\lfloor#1\rfloor}
\newcommand\ceil[1]{\lceil#1\rceil}
\newcommand\bceil[1]{\bigg\lceil#1\bigg\rceil}

\newcommand\ob[1]{\overline{\mathbf{#1}}}

\newcommand\nint[1]{\mathrm{nint}{#1}}

\newcommand\dx[2]{\frac{\mathrm{d} #1}{\mathrm{d} #2}}

\newcommand{\half}{\frac{1}{2}}

\definecolor{commentgreen}{RGB}{2,112,10}
\definecolor{eminence}{RGB}{108,48,130}
\definecolor{weborange}{RGB}{255,165,0}
\definecolor{frenchplum}{RGB}{129,20,83}
\definecolor{cbblue}{RGB}{5,113,176}
\definecolor{cbred}{RGB}{202,0,30}
\definecolor{cborange}{RGB}{244,165,130}

\definecolor{set1}{RGB}{228,26,28}
\definecolor{set2}{RGB}{55,126,184}
\definecolor{set3}{RGB}{77,175,74}
\definecolor{set4}{RGB}{152,78,163}
\definecolor{pale1}{RGB}{251,180,174}
\definecolor{pale2}{RGB}{179,205,227}
\definecolor{pale3}{RGB}{204,235,197}
\definecolor{pale4}{RGB}{222,203,228}

\newcommand\CONDITION[2]

\journal{}
\begin{document}

\begin{frontmatter}

\title{Inline Vector Compression for Computational Physics}

\author{W. Trojak\corref{cor1}~\orcid{0000-0002-4407-8956}}
\ead{wt247@tamu.edu}
\cortext[cor1]{Corresponding author}
\address{Department of Ocean Engineering, Texas A\&M University, College Station}

\author{F. D. Witherden\corref{cor2}~\orcid{0000-0003-2343-412X}}
\ead{fdw@tamu.edu}
\address{Department of Ocean Engineering, Texas A\&M University, College Station}

\begin{abstract}
A novel inline data compression method is presented for single-precision vectors in three dimensions. The primary application of the method is for accelerating computational physics calculations where the throughput is bound by memory bandwidth. The scheme employs spherical polar coordinates, angle quantisation, and a bespoke floating-point representation of the magnitude to achieve a fixed compression ratio of 1.5. The anisotropy of this method is considered, along with companding and fractional splitting techniques to improve the efficiency of the representation. We evaluate the scheme numerically within the context of high-order computational fluid dynamics. For both the isentropic convecting vortex and the Taylor--Green vortex test cases, the results are found to be comparable to those without compression. Performance is evaluated for a vector addition kernel on an NVIDIA Titan V GPU; it is demonstrated that a speedup of 1.5 can be achieved.
\end{abstract}


\end{frontmatter}



\section{Introduction}
\label{sec:intro}

Over the past 30 years, improvements in computing capabilities, measured in floating-point operations per second, have consistently outpaced improvements in memory bandwidth~\cite{Witherden2014}.  A consequence of this is that many numerical methods for partial differential equations are memory bandwidth bound algorithms. Additionally, over the past 10 years, there has been a trend towards the use of graphics processing units (GPUs) for simulations. These accelerators have increased computational power and memory bandwidth compared to CPUs. However, one limitation is that the total amount of memory is substantially less than what would accompany an equivalently capable CPU. To this end, there is substantial interest in techniques that can conserve either memory or memory bandwidth.

Arguably the simplest technique is to migrate the numerical scheme from eight-byte double-precision numbers to four-byte single-precision numbers. Such migration has the potential to halve the bandwidth and storage requirements for a code.  The feasibility of this depends heavily on both the numerical scheme and its application area.  Studies indicate that single-precision arithmetic may be appropriate in computational geosciences, molecular dynamics, and implicit large eddy simulation~\cite{Witherden2019,Homann2007,Bailey2005,Trojak2020}.  More recently, driven primarily by the requirements of the machine learning community, hardware vendors have started to include support for two-byte half-precision arithmetic~\cite{Nvidia2017}.  However, the restricted range of half-precision floating-point numbers~\cite{IEEE754_2008} renders them unsuitable for general-purpose computation.

Another promising approach is that of in-line data compression.  This takes advantage of the fact that the data in scientific simulations often exhibits spatial correlations and is hence amenable to compression.  When evaluating compression schemes in the context of scientific computing, there are several factors to consider. The first is whether the scheme is lossless or lossy.  Next is the extent to which the scheme admits random access to the compressed data.  A third factor is the degree of asymmetry, if any, between the cost of compression and that of decompression.  For schemes that are lossy, a fourth factor to consider is if the compression ratio is fixed or variable.  Additionally, with lossy schemes, a further consideration is the ability to bound the magnitude of any loss.

A variety of compression schemes, both general purpose and domain-specific, have been proposed.  When considering lossless compression, it follows from the Pigeonhole principle that no scheme can guarantee to reduce the size of an input.  An immediate consequence of this is that the compression ratio of a lossless scheme must always be a function of the input.  A corollary of this is that the maximum problem size in which a code employing lossless in-line compression can robustly handle is identical to that of the same code without compression.  Moreover, the variability in the compression ratio makes it difficult to predict the run-time performance of such a code.  As such, within the context of in-line compression for scientific codes, lossy schemes are preferable.

The primary advantage of lossy compression schemes is that not only can higher compression ratios be obtained, but they may admit a fixed compression ratio.  Some  early work on block data compression was aimed at computer graphics and was based on Vector Quantisation (VQ)~\cite{Delp1979,Ning1992,Campbell1986}. A prominent methodology within VQ is the LBG approach of Linde~\etal~\cite{Linde1980}. This approach used a compression codebook learned \textit{a priori} through optimisation. This allows for efficient vector decompression with a low error when the vector domain is similar to the learned codebook sub-domain. Subsequent methods that have evolved from this include the S3TC method~\cite{Iourcha1999}, which operates on textures consisting of several pixels. These techniques typically have a large asymmetry between compression and decompression, with compression being orders of magnitude slower. An additional method of importance is the adaptation of LBG by Schneider~\etal~\cite{Schneider2003}, which employed a codebook initialisation method using principal component analysis based splitting. 

More recently, the ZFP algorithm and library were developed~\cite{Lindstrom2014a} to work across small blocks ($4^d$ in size, where $d$ is the dimensionality) with user-controlled fixed-rate compression. This method allowed for independent block access and boasted near symmetric compression and decompression. However, due to the larger block size, this is inappropriate for our desired unstructured application. An algorithm more closely related to the issue of memory-bound compute is the BLOSC methodology and library~\cite{Alted2010}. BLOSC utilises several compression techniques, including DEFLATE and related methods as well as LZ compression~\cite{Ziv1977,Ziv1978}. LZ compression is designed for compression of sequential data-sets, and as a consequence, BLOSC is highly efficient for bandwidth limited calculations with sequential or blocked data access. Yet, similar to ZFP, these techniques are not well suited to the unstructured data access we are concerned with.
	
Evidently, if compression is to improve the performance of a numerical scheme, a baseline requirement is that the decompression rate must exceed memory bandwidth.  An associated requirement is that compression and decompression have comparable costs. This is to enable the compression of dynamic data; that is to say, the output generated by kernels rather than just constant inputs.  A tertiary requirement, albeit one which is particularly relevant for numerical schemes on unstructured grids, is for there to be little to no penalty associated with random access to data. Finally, it is desired that the multiple applications of compression/decompression to the same data should not result in substantially different data, i.e. the compression/decompression operator should be idempotent or near-idempotent. 
	
A more applicable set of techniques arises from the compression of individual three-component pixel data. A review of some techniques in this area was presented by Cigolle~\etal~\cite{Cigolle2014}. It has also been shown by Meyer~\etal~\cite{Meyer2010} that 51-bits is sufficient to losslessly represent unit vectors formed of three 32-bit floating-point numbers. More recently,  Smith~\etal~\cite{Smith2012a} considered applying lossy compression to this case. The approach quantizes three component unit vectors by initially transforming them into spherical coordinates, and subsequently discretising the two angles into bins.  In this approach the number of bins depends on both the desired error tolerance and the angles themselves.  Using this approach it is possible to reduce the storage requirement from 96-bits to between eight and 30-bits depending on the desired error tolerance and the inputs themselves.  Since the scheme generates output at the bit level and admits a variable compression ratio it is only suitable for applications whose access patterns are both serial and sequential.

In this paper, we propose a scheme for the compression of triplets of single-precision numbers which occur frequently in computational physics applications.  Building upon the ideas presented in  Smith~\etal~\cite{Smith2012a} this approach is capable of compressing three component vectors of \emph{any} magnitude whilst achieving a fixed $50\%$ compression ratio---reducing a 96-bit vector down to 64-bits.  This ratio enables non-sequential and high-performance parallel access to vectors, even on hardware platforms with eight byte alignment requirements.  Despite the fact that this scheme has to deal with the additional complexities associated with non-unit vectors the overall compression and decompression routines end up being somewhat less computationally expensive than those of Smith~\etal. Furthermore, complexity of the compression and decompression routines in our proposed approach is similar, enabling it to be employed in an in-line capacity.

The remainder of this paper is structured as follows.  In Section~\ref{sec:comp}, we outline our compression scheme and analyse its numerical properties through a series of synthetic benchmarks.  Implementation details are discussed in Section~\ref{sec:impl}.  In Section~\ref{sec:numerics}, we implement our scheme into a high-order computational fluid dynamics code and assess its impact on accuracy. Here the performance of our approach on an NVIDIA Titan V GPU is evaluated.  Finally, the conclusions are discussed in Section~\ref{sec:conclusions}.
\section{Methodology}
\label{sec:comp}

\subsection{Basic Compression Methodology}\label{ssec:basic}
   Given a vector $\mathbf{x} = [x,y,z]^T$, we begin by writing the transformation from Cartesian to spherical polar coordinates as
	\begin{equation}\label{eq:cart2sphere}
		\mathbf{x} = \begin{bmatrix} x \\ y \\ z \end{bmatrix} =
		\begin{bmatrix} r\cos{\theta}\sin{\phi} \\ r\sin{\theta}\sin{\phi} \\ r\cos{\phi} \end{bmatrix},
	\end{equation}
	where $r$, $\theta$ and $\phi$ are defined as
	\begin{subequations}\label{eq:spherical}
		\begin{align}
			r =& \|\mathbf{x}\|_2, \\
			\theta =& \tan^{-1}\bigg(\frac{y}{x}\bigg), \\
			\phi =& \cos^{-1}\bigg(\frac{z}{r}\bigg).
		\end{align}
	\end{subequations}
	Here $\phi \in [0,\pi)$ and $\theta\in[-\pi,\pi)$.  We note here that both of these ranges are substantially smaller than those which can be represented by a 32-bit floating-point number.  As a starting point, we consider discretising these angles by representing them as a pair of integers according to
	\begin{equation}\label{eq:a_disc}
			n_\theta = \nint{\bigg(\frac{n_{\theta\:\mathrm{max}}(\theta + \pi)}{2\pi}\bigg)} \quad \text{and} \quad
			n_\phi = \nint{\bigg(\frac{n_{\phi\:\mathrm{max}}\phi}{\pi}\bigg)},
	\end{equation}
	where $n_{\theta\:\mathrm{max}}$ and $n_{\phi\:\mathrm{max}}$ are defined as the maximum values supported by the integer type used to store $\theta$ and $\phi$, respectively. The function $\nint{(x)}$ is defined as that which rounds $x$ to the nearest  integer as per
	\begin{equation}
		\nint{(x)} = \bceil{\frac{\floor{2x}}{2}},
	\end{equation}
	where we have adopted the \emph{round half up} tie-breaking rule.  The corresponding inverse mappings are
	\begin{equation}\label{eq:a_inv}
			\hat{\theta} = \pi\bigg(\frac{2n_\theta}{n_{\theta\:\mathrm{max}}} - 1\bigg) \quad \text{and} \quad
			\hat{\phi} = \bigg(\frac{\pi n_\phi}{n_{\phi\:\mathrm{max}}}\bigg).
	\end{equation}
	Here, the hat is used to represent a reconstructed angle. The associated absolute errors are then given according to
	\begin{equation}\label{eq:error_def}
			\theta = \hat{\theta} + \epsilon_\theta +\eta_\theta \quad \text{and} \quad
			\phi = \hat{\phi} + \epsilon_\phi + \eta_\phi,
	\end{equation}
	where $\epsilon$ is a quantisation error and $\eta$ is an arithmetic error.  To proceed we shall assume that it is possible to implement the aforementioned compression procedure in a numerically stable manner such the arithmetic error is insignificant compared to the quantisation error. This point shall be explored in Section~\ref{sec:impl}. From the definitions of $\theta$ and $\phi$ it is clear that
	\begin{equation}\label{eq:abs_e}
			\epsilon_\theta \in \frac{\pi}{n_{\theta\:\mathrm{max}}}[-1,1] \quad \text{and} \quad
			\epsilon_\phi \in \frac{1}{2}\frac{\pi}{n_{\phi\:\mathrm{max}}}[-1,1].
	\end{equation}
	The quantisation errors on $\theta$ and $\phi$ then give rise to a reconstructed vector $\hat{\mathbf{x}} = [\hat{x},\hat{y},\hat{z}]^T$ with:
	\begin{subequations}
		\begin{align}
			\hat{x} &= r\cos{(\theta +\epsilon_\theta)}\sin{(\phi +\epsilon_\phi)}, \\
			\hat{y} &= r\sin{(\theta +\epsilon_\theta)}\sin{(\phi +\epsilon_\phi)},\\
			\hat{z} &= r\cos{(\phi +\epsilon_\phi)}.
		\end{align}
	\end{subequations}
	Let us define then error vector as $[\epsilon_x,\epsilon_y,\epsilon_z]^T = \hat{\mathbf{x}} - \mathbf{x}$.  Applying the trigonometric sum-to-product identities and assuming $(|\epsilon_\theta|,|\epsilon_\phi|) \ll 1$ we find
	\begin{subequations}\label{eq:v_error}
		\begin{align}
			\epsilon_x &\approx r(\epsilon_\phi\cos{\theta}\cos{\phi} -\epsilon_\theta\sin{\theta}\sin{\phi}), \\
			\epsilon_y &\approx r(\epsilon_\phi\sin{\theta}\cos{\phi} +\epsilon_\theta\cos{\theta}\sin{\phi}),\\
			\epsilon_z &\approx r\epsilon_\phi\sin{\phi}.
		\end{align}
	\end{subequations}
		
	\subsection{Base Scheme Evaluation}\label{ssec:base_eval}
\begin{figure}[tbhp]
		\centering
			\resizebox{0.8\linewidth}{!}{\input{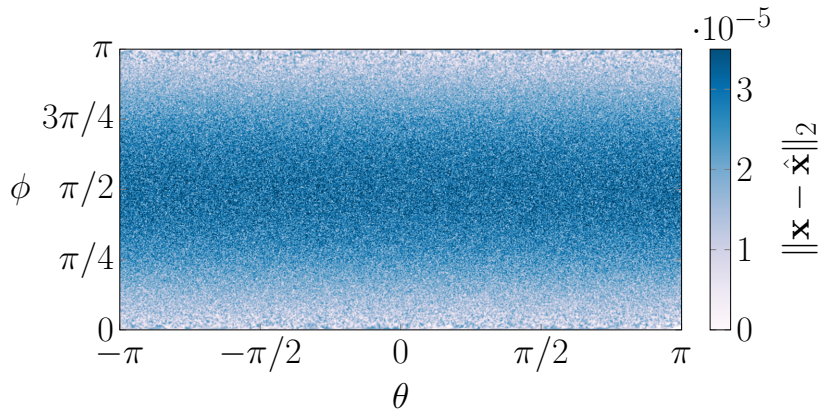}}
		\caption{Base scheme compression error on unit ball.}
		\label{fig:e_1616}
	\end{figure}

	We now wish to evaluate the error of the base method by which we mean discretising $n_\theta$ and $n_\phi$ with 16-bits and storing the magnitude as a standard single-precision floating-point (SPFP). This evaluation was performed by randomly sampling $\theta$ and $\phi$ such that the Cartesian points are uniformly distributed over a unit sphere. This, together with the computation of $\mathbf{x}$, is carried out using double-precision floating-point (DPFP) numbers. Then, before $\mathbf{x}$ is compressed, the components are converted to SPFP. This was to ensure that all points on the unit ball, in single-precision, could be sampled. To  minimise the arithmetic error in the compression and decompression steps, all the intermediary real numbers used were cast as DPFP numbers. Then the $L_2$ norm of the error between the SPFP $\mathbf{x}$ and the vector after compression and decompression, $\hat{\mathbf{x}}$, was calculated. These results can be seen in Fig.~\ref{fig:e_1616}. Here, the level of anisotropy in the method is clearly visible, with both the error described in Eq.~(\ref{eq:v_error}) and the larger range of $\theta$ contributing to this. 
	
	\subsection{Increasing Discretisation Resolution}\label{ssec:bits}
		In Sections~\ref{ssec:basic} and \ref{ssec:base_eval}, the basic compression methodology was presented and evaluated. We now wish to investigate if the error magnitude and degree of anisotropy can be reduced while still limiting the result to 64-bits. It is proposed that bits from the vector magnitude may be redistributed to increase the quantisation on the angles. An aspect of this compression technique that should be used to inform bit redistribution is that the range of $\theta$  is twice that of $\phi$. From Eq.~(\ref{eq:abs_e}~\&~\ref{eq:v_error}), it can be seen that this will impact the error; hence, $\theta$ should be quantised with an additional bit.
		
		The most straightforward bit to repurpose is the sign bit of $r$ since the magnitude will always be positive. Two subsequent techniques will be introduced as a means of recovering further bits from the magnitude. The first method is to remove bits from the exponent of the SPFP number representing $r$. This is motivated by the maximum and minimum values of the exponent being $2^{127}\approx 10^{38}$ and $2^{-126}\approx10^{-38}$, respectively. This range of values is larger than is typically encountered in the simulation of physical phenomena, and hence it may be possible to reduce the range of the exponent without adversely affecting the results. Such a reduction may be accomplished as follows, beginning by describing the procedure for calculating the exponent~\cite{IEEE754_2008}, $e$, as
		\begin{equation}
			e = E_8 - b_8,
		\end{equation}
		where $E_8$ is the $8$ bit integer representation of the exponent bits stored in the SPFP number and $b_8$ is a bias that shifts the range of $E_8$ such that $e$ can take negative values. In the case of SPFP, $b_8=127$ such that $e\in\{-126,\dots, 127\}$.
				
		Reducing the exponent length, instead storing $E_7$, will then lead to a reduction in the range of $e$. Therefore, the method is
		\begin{equation}
			\hat{E}_8 = E_7 + b_8 - b_7,
		\end{equation}
		where $b_7$ is a new bias to be set and $\hat{E}_8$ is the reconstructed exponent integer needed for SPFP arithmetic. It can then be seen that this will lead to
		\begin{equation}
			e = \hat{E}_8 - b_8 = E_7 - b_7, 
		\end{equation}				
		such that
		\[
		-(b_7+1) \le e \le 2^7-(b_7+2).
		\]
		When removing a single bit from the exponent it was hypothesised that $b_7=80$ was a reasonable compromise for computational fluids calculations. An asymmetric bias was chosen based on the hypothesis that, often in physical simulations, small numbers and zeros are important in the phenomena exhibited. Conversely, very large numbers in properly normalised calculations are typically indicative of a diverging solution and hence limiting the range here is deemed reasonable. However, the degree of exponent compression and bias could be varied on a case-by-case basis.
		
		Going further, we will now consider how bits may be removed from the mantissa. Starting from SPFP numbers~\cite{IEEE754_2008}, the mantissa, $T$, or the fraction, is defined using 24-bits as:
		\begin{equation}
			T_{24} = d_0.d_1\cdots d_{22}d_{23} = 1.d_1\cdots d_{22}d_{23},
		\end{equation}
		where, through proper normalisation, $d_0$ is assumed to always be $1$ and hence only $23$ bits are stored. Bits may then be removed from the tail of the mantissa and repurposed.  Taking the example of removing one bit, $d_{23}$, and replacing it by zero in the reconstructed mantissa we get
		\begin{equation}
			\hat{T}_{24} = 1.d_1\cdots d_{22}.
		\end{equation}
		
		Within the set of SPFP numbers there is a subset called normal numbers. This is the case when the bit $d_0$ is always $1$, i.e.\ the binary point of the mantissa is shifted such that the $d_0=1$. To normalise the number, the binary point is moved by increasing or decreasing the exponent. In the case of SPFP, if the exponent, $E_8$, equals 1 and $d_0\neq1$ then the number is said to be subnormal. The handling of this exception is usually done by the hardware based on the compiler options. However, with $E_7$ this has to be handled by the compression algorithm. Hence, we impose that subnormal numbers, i.e.\ when $E_7=1$, are set to the lowest supported value. 				
		
\begin{figure}[tbhp]
			\centering
			\begin{subfigure}[h]{0.8\linewidth}
				\centering
				\resizebox{\linewidth}{!}{\begin{tikzpicture}[node distance=0pt, every node/.style={outer sep=0, text height=4ex,text depth=0.25ex}]
  	\coordinate (spe) at (0,0);
  	\coordinate (spm) at (4,0);
  	\coordinate (spp) at (15.5,0);
  	\coordinate (spt) at (23.5,0);
  	\coordinate (spf) at (32,0);
  	\coordinate (t) at (0,1.25);
	
	\node at (0.25,-2) (E) {\Huge 63};	
	\draw [ultra thick, -*, draw=black] (E) -- (0.25,-0.25);
	\path [ultra thick, draw=black, *-*] (0.1,1.5) -- node [midway,above=0.8em] {\Huge Exponent} (3.9,1.5);
	\draw [ultra thick, draw=black, fill={Pastel1-A}]  (spe) grid[xstep=0.5,ystep=1.25](spm |- t) rectangle (spe);
	\node at (3.75,-2) (E) {\Huge 56};	
	\draw [ultra thick, -*, draw=black] (E) -- (3.75,-0.25);
	
	\path [ultra thick, draw=black, *-*] (4.1,1.5) -- node [midway,above=0.8em] {\Huge Mantissa} (15.4,1.5);
	\draw [ultra thick, draw=black, fill={Pastel1-B}] (spm) grid[xstep=0.5,ystep=1.25](spp |- t) rectangle (spm);
	\node at (15.25,-2) (M) {\Huge 33};	
	\draw [ultra thick, -*, draw=black] (M) -- (15.25,-0.25);
	
	\path [ultra thick, draw=black, *-*] (15.6,1.5) -- node [midway,above=0.8em] {\Huge $n_\phi$} (23.4,1.5);
	\draw [ultra thick, draw=black, fill={Pastel1-C}] (spp) grid[xstep=0.5,ystep=1.25](spt |- t) rectangle (spp);
	\node at (23.25,-2) (E) {\Huge 17};	
	\draw [ultra thick, -*, draw=black] (E) -- (23.25,-0.25);
	
	\path [ultra thick, draw=black, *-*] (23.6,1.5) -- node [midway,above=0.8em] {\Huge $n_\theta$} (31.9,1.5);
	\draw [ultra thick, draw=black, fill={Pastel1-D}] (spt) grid[xstep=0.5,ystep=1.25](spf |- t) rectangle (spt);
	\node at (31.75,-2) (E) {\Huge 0};	
	\draw [ultra thick, -*, draw=black] (E) -- (31.75,-0.25);

\end{tikzpicture}}
				\caption{$\langle0,8,23\rangle\text{-}16\text{-}17$}
				\label{fig:signbit}
			\end{subfigure}
			\\
			\begin{subfigure}[h]{0.8\linewidth}
				\centering
				\resizebox{\linewidth}{!}{\begin{tikzpicture}[node distance=0pt, every node/.style={outer sep=0, text height=4ex,text depth=0.25ex}]
	\coordinate (spe) at (0,0);
  	\coordinate (spm) at (3.5,0);
  	\coordinate (spp) at (15,0);
  	\coordinate (spt) at (23.5,0);
  	\coordinate (spf) at (32,0);
  	\coordinate (t) at (0,1.25);
	
	\node at (0.25,-2) (E) {\Huge 63};	
	\draw [ultra thick, -*, draw=black] (E) -- (0.25,-0.25);
	\path [ultra thick, draw=black, *-*] (0.1,1.5) -- node [midway,above=0.8em] {\Huge Exponent} (3.4,1.5);
	\draw [ultra thick, draw=black, fill={Pastel1-A}]  (spe) grid[xstep=0.5,ystep=1.25](spm |- t) rectangle (spe);
	\node at (3.25,-2) (E) {\Huge 57};	
	\draw [ultra thick, -*, draw=black] (E) -- (3.25,-0.25);
	
	\path [ultra thick, draw=black, *-*] (3.6,1.5) -- node [midway,above=0.8em] {\Huge Mantissa} (14.9,1.5);
	\draw [ultra thick, draw=black, fill={Pastel1-B}] (spm) grid[xstep=0.5,ystep=1.25](spp |- t) rectangle (spm);
	\node at (14.75,-2) (M) {\Huge 34};	
	\draw [ultra thick, -*, draw=black] (M) -- (14.75,-0.25);
	
	\path [ultra thick, draw=black, *-*] (15.1,1.5) -- node [midway,above=0.8em] {\Huge $n_\phi$} (23.4,1.5);
	\draw [ultra thick, draw=black, fill={Pastel1-C}] (spp) grid[xstep=0.5,ystep=1.25](spt |- t) rectangle (spp);
	\node at (23.25,-2) (E) {\Huge 17};	
	\draw [ultra thick, -*, draw=black] (E) -- (23.25,-0.25);
	
	\path [ultra thick, draw=black, *-*] (23.6,1.5) -- node [midway,above=0.8em] {\Huge $n_\theta$} (31.9,1.5);
	\draw [ultra thick, draw=black, fill={Pastel1-D}] (spt) grid[xstep=0.5,ystep=1.25](spf |- t) rectangle (spt);
	\node at (31.75,-2) (E) {\Huge 0};	
	\draw [ultra thick, -*, draw=black] (E) -- (31.75,-0.25);
\end{tikzpicture}}
				\caption{$\langle0,7,23\rangle\text{-}17\text{-}17$}
				\label{fig:expbit}
			\end{subfigure}
			\\
			\begin{subfigure}[h]{0.8\linewidth}
				\centering
				\resizebox{\linewidth}{!}{\begin{tikzpicture}[node distance=0pt, every node/.style={outer sep=0, text height=4ex,text depth=0.25ex}]
	\coordinate (spe) at (0,0);
  	\coordinate (spm) at (3.5,0);
  	\coordinate (spp) at (14.5,0);
  	\coordinate (spt) at (23,0);
  	\coordinate (spf) at (32,0);
  	\coordinate (t) at (0,1.25);
	
	\node at (0.25,-2) (E) {\Huge 63};	
	\draw [ultra thick, -*, draw=black] (E) -- (0.25,-0.25);
	\path [ultra thick, draw=black, *-*] (0.1,1.5) -- node [midway,above=0.8em] {\Huge Exponent} (3.4,1.5);
	\draw [ultra thick, draw=black, fill={Pastel1-A}]  (spe) grid[xstep=0.5,ystep=1.25](spm |- t) rectangle (spe);
	\node at (3.25,-2) (E) {\Huge 57};	
	\draw [ultra thick, -*, draw=black] (E) -- (3.25,-0.25);
	
	\path [ultra thick, draw=black, *-*] (3.6,1.5) -- node [midway,above=0.8em] {\Huge Mantissa} (14.4,1.5);
	\draw [ultra thick, draw=black, fill={Pastel1-B}] (spm) grid[xstep=0.5,ystep=1.25](spp |- t) rectangle (spm);
	\node at (14.25,-2) (M) {\Huge 35};	
	\draw [ultra thick, -*, draw=black] (M) -- (14.25,-0.25);
	
	\path [ultra thick, draw=black, *-*] (14.6,1.5) -- node [midway,above=0.8em] {\Huge $n_\phi$} (22.9,1.5);
	\draw [ultra thick, draw=black, fill={Pastel1-C}] (spp) grid[xstep=0.5,ystep=1.25](spt |- t) rectangle (spp);
	\node at (22.75,-2) (E) {\Huge 18};	
	\draw [ultra thick, -*, draw=black] (E) -- (22.75,-0.25);
	
	\path [ultra thick, draw=black, *-*] (23.1,1.5) -- node [midway,above=0.8em] {\Huge $n_\theta$} (31.9,1.5);
	\draw [ultra thick, draw=black, fill={Pastel1-D}] (spt) grid[xstep=0.5,ystep=1.25](spf |- t) rectangle (spt);
	\node at (31.75,-2) (E) {\Huge 0};	
	\draw [ultra thick, -*, draw=black] (E) -- (31.75,-0.25);
\end{tikzpicture}}
				\caption{$\langle0,7,22\rangle\text{-}17\text{-}18$}
				\label{fig:manbit}
			\end{subfigure}
			\caption{Compression regimes under consideration.}
			\label{fig:comp_regimes}
		\end{figure}
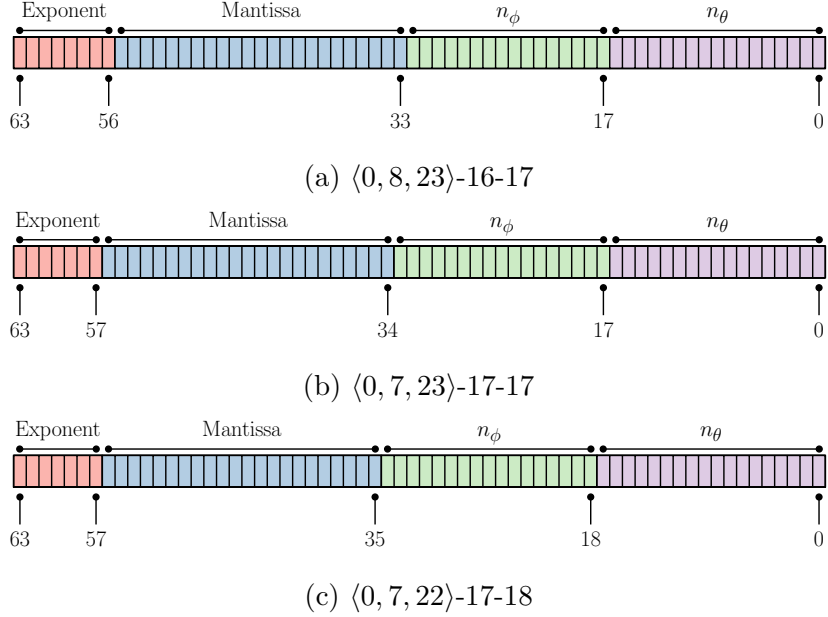

	Throughout the remainder of this work we will use the following notation to describe how the 64 bits are used: $\langle s,e,m\rangle\text{-}p\text{-}t$, where $s$ indicates the presence of a sign bit; $e$ is the number of exponent bits; $m$ is the number of mantissa bits; and $p$ and $t$ are the number of bits in $n_\phi$ and $n_\theta$ respectively. Some potential bit layouts are shown in Fig.~\ref{fig:comp_regimes} which are investigated later.
		
		\begin{figure}[tbhp]
			\centering
			\begin{subfigure}[h]{0.8\linewidth}
				\centering
					\resizebox{\linewidth}{!}{\input{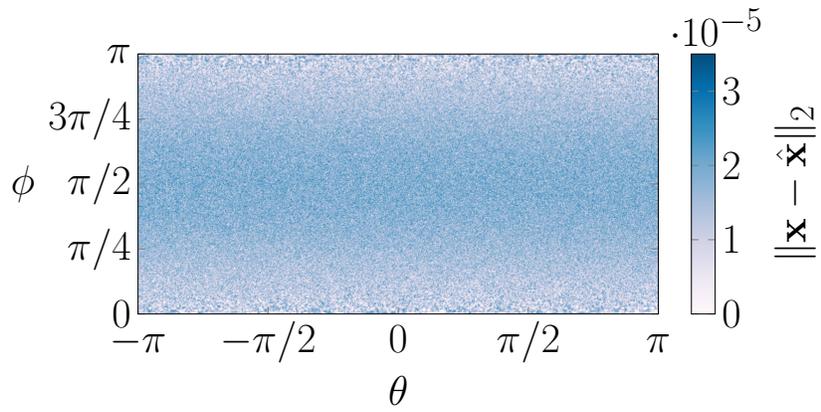}}
				\caption{$\langle0,8,23\rangle\text{-}16\text{-}17$.}
				\label{fig:e_1617}
			\end{subfigure}
			\\
			\begin{subfigure}[h]{0.8\linewidth}
				\centering
					\resizebox{\linewidth}{!}{\input{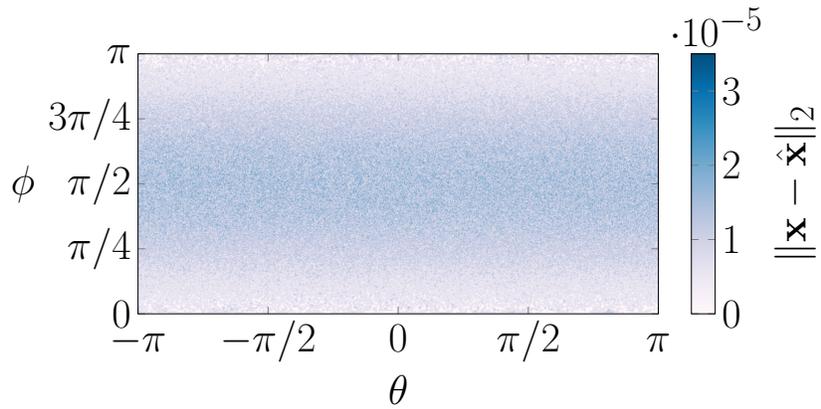}}
				\caption{$\langle0,7,23\rangle\text{-}17\text{-}17$.}
				\label{fig:e_1717}
			\end{subfigure}
			\\
			\begin{subfigure}[h]{0.8\linewidth}
				\centering
				\resizebox{\linewidth}{!}{\input{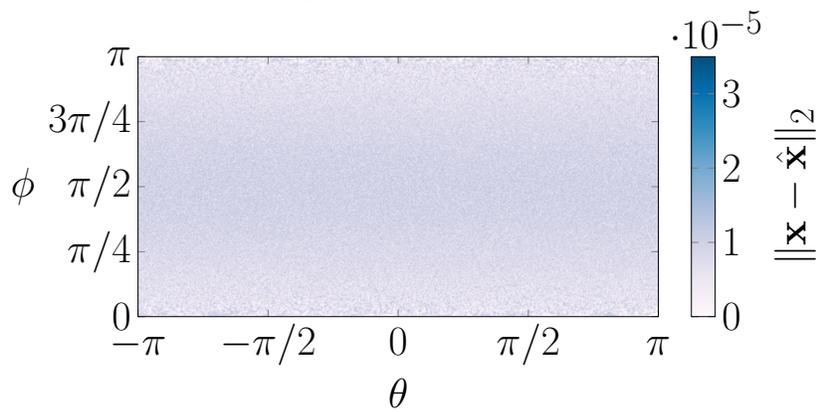}}
				\caption{$\langle0,7,22\rangle\text{-}17\text{-}18$.}
				\label{fig:e_1718}
			\end{subfigure}
			\caption{Error of random vector field on unit ball for $10^8$ samples.}
			\label{fig:error}
		\end{figure}
	
	We will now compare the error on the unit ball of three different bit distributions: $\langle0,8,23\rangle\text{-}16\text{-}17$,  $\langle0,7,23\rangle\text{-}17\text{-}17$, and $\langle0,7,22\rangle\text{-}17\text{-}18$.	The methodology followed here is the same as in Section~\ref{ssec:basic}. The impact of bit utilisation on the error is shown in Fig.~\ref{fig:error}. It is clear that, not only are all three methods an improvement over the base scheme shown in Fig.~\ref{fig:e_1616}, but that the additional angle resolution has a significant impact on the error. In this instance, the average error over the samples was found to be $1.55\times10^{-5}$, $1.07\times10^{-5}$, and $7.74\times10^{-6}$, respectively. 
	
	A further investigation was performed where the average error was calculated while varying the ball radius in $[10^{-8},10^8]$. It was found that the average error, normalised by the radius, was approximately constant. Therefore, in this range, the exponent and mantissa contraction has not led to any additional dependence of error on $r$.  
	
	\subsection{Fractional Splitting}
	The schemes so far described have discretised the angles using powers of two due to their connection to binary representations. For example, $n_{\phi,\mathrm{max}}=2^{16}-1$, $n_{\theta,\mathrm{max}}=2^{17}-1$. It is proposed that it may be possible to split the available discretisation at some other interface. This can be defined in the general form for some number $n_p$, where the total number of bits available is $p$, and $n_\phi$ and $n_\theta$ are encoded based on:
	\begin{subequations}\label{eq:q_split}
		\begin{align}
			n_p =&\: n_\phi (n_{\theta,\mathrm{max}}+1) + n_\theta, \\
		    n_{p,\mathrm{max}} =&\: (n_{\phi,\mathrm{max}}+1)(n_{\theta,\mathrm{max}}+1) -1 < 2^p.
		\end{align}
	\end{subequations}
	For the case of $p=35$, where the additional bits have been obtained through a reduction in the exponent range and mantissa precision, we will vary $n_{\phi, \mathrm{max}}+1 \in [2^{16},\ldots, 2^{18}]$. By setting $n_{\phi,\mathrm{max}}$, Eq.~(\ref{eq:q_split}) then constrains $n_{\theta,\mathrm{max}}$. For each of the splitting choices we will then calculate the mean and standard sample deviation of the error on the unit ball, as defined by
	\begin{equation}\label{eq:stat}
		e_i = \|\mathbf{x}_i - \hat{\mathbf{x}}_i\|_2, \quad 
		\overline{e} = \frac{1}{N}\sum_{i=1}^Ne_i, \quad
		\sigma(e)^2 = \frac{1}{N-1}\sum^N_{i=1}(e_i-\overline{e})^2.
	\end{equation}

	\begin{figure}[tbhp]
		\centering
		\begin{subfigure}[h]{0.48\linewidth}
			\centering
				\resizebox{\linewidth}{!}{	\begin{tikzpicture}
		\begin{axis}[name=plot1,xlabel={$n_{\phi,\mathrm{max}}$},ylabel={$\overline{e}$},
		    xtick={65536,131072,262144},ytick={0.000007,0.000009,0.000011,0.000013},
		    xticklabels={$2^{16}$,$2^{17}$,$2^{18}$},
    		xmin=65536,xmax=262144,
    		xmode=log,
    		log ticks with fixed point,
    		ymin=0.7e-5,ymax=1.3e-5,
    		style={font=\large}]
			\addplot[color={RdBu-C}, style={thick}]
				table[x=np,y=mu,col sep=comma,unbounded coords=jump]{./Figs/data/r1_frac_sd.csv};
		\end{axis} 		
	\end{tikzpicture}}
			\caption{Mean error.}
			\label{fig:ebar_1718}
		\end{subfigure}
		~
		\begin{subfigure}[h]{0.48\linewidth}
			\centering
				\resizebox{\linewidth}{!}{	\begin{tikzpicture}
		\begin{axis}[name=plot1,xlabel={$n_{\phi,\mathrm{max}}$},ylabel={$\sigma(e)$},
		    xtick={65536,131072,262144},
		    ytick={0.000003,0.000004,0.000005,0.000006,0.000007},
		    xticklabels={$2^{16}$,$2^{17}$,$2^{18}$},
    		xmin=65536,xmax=262144,
		    xmode=log,
    		log ticks with fixed point,
    		ymin=0.3e-5,ymax=0.7e-5,
    		style={font=\large}]
			\addplot[color={RdBu-M}, style={thick}]
				table[x=np,y=sigma,col sep=comma,unbounded coords=jump]{./Figs/data/r1_frac_sd.csv};
		\end{axis} 		
	\end{tikzpicture}}
			\caption{Error standard deviation.}
			\label{fig:sd_1718}
		\end{subfigure}
		\caption{Fractional splitting error for a 35-bit quantisation.  Plots were obtained with $N=10^4$ samples.}
		\label{fig:frac_1718}
	\end{figure}
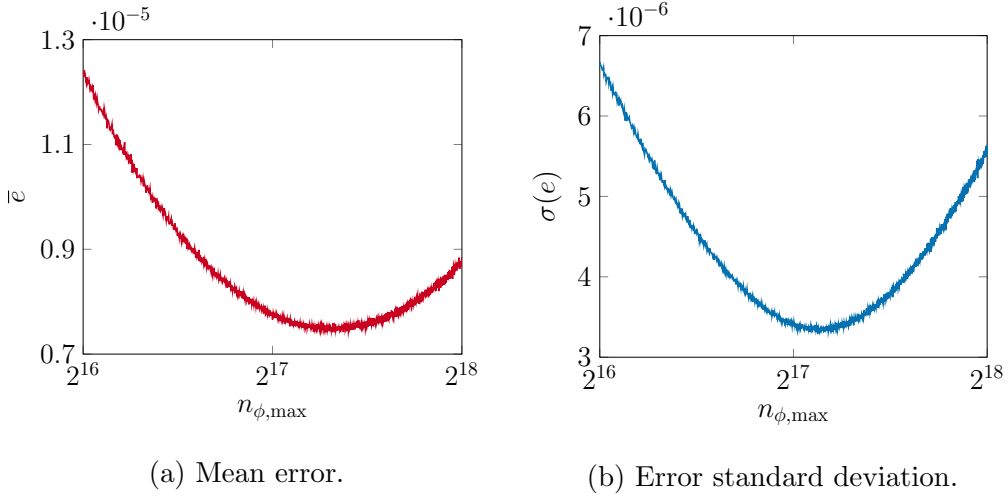
		
	As evident from Fig.~\ref{fig:frac_1718}, variation of the splitting location from $n_{\phi,\mathrm{max}}=2^{17}$ has only a minor impact on the error exhibited by the compression method. Therefore, as the  reduction in $\sigma$ that can be achieved is small and as fractional splitting would increase code complexity, it is deemed to be of insufficient benefit to be used. 
	 
\subsection{Angle Discretisation Companding}\label{ssec:companding}
	Equation~(\ref{eq:v_error}), together with Eq.~(\ref{eq:abs_e}), made clear that a uniform discretisation will give rise to anisotropy in this compression method. We will now discuss some alternative discretisation methods with the aim of reducing this anisotropy. Firstly, for an angle discretisation to be practical it must be easily invertible; for example, Eq.~(\ref{eq:a_disc}) is straightforwardly inverted as in Eq.~(\ref{eq:a_inv}). Therefore, we can straightforwardly use any transformation where $n=\nint{f(\psi)}$ and $\hat{\psi}=f^{-1}(n)$ are computable in finite time. This does, however, severely limit the possible choices to functions implemented—together with their inverse—by compilers. For example, $n_\mathrm{max}f(\psi) = \psi$ is admissible but $n_\mathrm{max}f(\psi) = \psi + \sin{(4\pi\psi)}/4\pi$ is not. This latter example was chosen as, from the previous error analysis, it can be thought to be a good candidate to reduce the error. However, to invert the discretisation, a root finding method, such as the fixed point iteration, has to be used. This is due to the stationary points in the mapping which can lead to large inaccuracies when a high degree of angle discretisation is used. That leads us to the question that, if stationary points could be avoided, could it be permissible to use a mapping that is not simply inverted? At the present time it is believed that it may be possible, but the root finding method is likely to be computationally expensive, especially to the desired level of precision, which in this case is close to machine precision. 
	
	Smith~\etal~\cite{Smith2012a} proposed a method where the number of bins used for $\theta$ was varied based on an error tolerance, $\tau$, and the value of $\phi$ that would remove anisotropy. This was defined as
	\begin{subequations}
		\begin{align}
			\frac{1}{\hat{n}_{\theta,\mathrm{max}}} =&\: \cos^{-1}{\bigg(\frac{\cos{\tau} - \cos{\phi}\cos{\big(\phi + \frac{\pi}{2n_{\phi,\mathrm{max}}} \big)}}{\sin{\phi}\sin{\big(\phi + \frac{\pi}{2n_{\phi,\mathrm{max}}}\big)}}\bigg)}, \\
			n_{\theta,\mathrm{max}} =&\: \ceil{\pi\hat{n}_{\theta,\mathrm{max}}}.
		\end{align}
	\end{subequations}		

	From this definition it should be clear that this poses a high computational cost for both compression and decompression, due to several trigonometric function evaluations, each of which can run to hundreds of clock cycles. Furthermore, it can be shown that for this relationship, the maximum number of bins as the tolerance tends to zero is equal to $2n_{\phi,\mathrm{max}}$, which in turn means that this method achieves isotropy by removing $\theta$ bins as $\phi\rightarrow\pi/2$. Therefore, resolution will be wasted when fixed rate compression is used. Hence, this method is deemed unsuitable for the application proposed here.
		
Two alternative invertible mappings were tested. The first was $n=\nint{(n_\mathrm{max}(1-\cos{\psi})/2)}$, which does contain stationary points, and was found to give far worse error performance when applied to $\phi$ and $\theta$. A second mapping that was attempted was
	\begin{equation}
		n = \nint{\bigg(mn_\mathrm{max}\Big[\tanh{\big(\gamma(2\psi-1)\big)+c\Big]}\bigg)}, \quad c=\tanh{\gamma}, \quad m=\frac{1}{2c}.
	\end{equation}
	
Performance was again measured by the mean and standard deviation of the error defined in Eq.~(\ref{eq:stat}). It was found that $\gamma=0.5$ could only give a reduction in $\sigma(e)$ of $\sim2\%$. Therefore, it is clear that the additional cost of this discretisation distribution outweighs the benefits. A final attempt is made to reduce the  anisotropy by optimising the discretisation brackets to reduce $\sigma(e)$. This was initially performed for $n_{\theta,\mathrm{max}}=200$ and $n_{\phi,\mathrm{max}}=100$, with the error evaluated on a unit sphere at $\sim3\times10^{5}$ points. The optimisation was performed using a simulated annealing approach~\cite{Bertsimas1993} where the cost function was $\sigma(e)$. In this case, the uniform distribution gave $\sigma(e)\approx 7.3\times10^{-3}$ and the optimal distribution gave $\sigma(e)\approx4.8\times10^{-3}$. However, the optimal distribution found was non-smooth, with large variation in bucket width even when smoothing was applied.  It follows that the incorporation of such a distribution is likely to require a device such as a lookup table, which in turn has a negative impact on computational efficiency.
\section{Implementation}
\label{sec:impl}

    The analysis in the previous sections concluded that the average error and the absolute level of anisotropy can be reduced significantly when using a bit layout of $\langle0,7,22\rangle\text{-}17\text{-}18$. During these experiments, all intermediate calculations were performed using double-precision floating-point (DPFP) arithmetic. This has a clear increase in computational cost and, after characterising the schemes, we wished to move some or all of the calculation to single-precision.  However, through experimentation, differences in the error were noticed as the implementation of the compression routine was varied slightly. This highlighted that under some circumstances the assumption that arithmetic errors, $\eta$, were small is false unless care is taken during implementation of these methods. Here we present an exploration of these variations and demonstrate some of the sensitivities of the compression method to the intermediate working precision. Recommendations will be made based on this investigation as to how the arithmetic error can be kept small when implementing this technique. 
    
    In summary, three SPFP numbers, $\mathbf{x}$, are passed to a function where $\theta$, $\phi$, and $r$ are calculated according to Eq.~(\ref{eq:spherical}). Then, $\theta$ and $\phi$ are quantised and the exponent and mantissa of $r$ are manipulated. Finally, the resulting bits are combined and 64 bits are returned from the function. The effect of manipulating $r$ and the degree of quantisation of $\theta$ and $\phi$ were explored in Section~\ref{sec:comp}, but the intermediate precision when calculating and quantising $\theta$ and $\phi$ were not discussed. 
    
    When calculating $\theta$ or $\phi$ there are two clear options: all intermediates are either SPFP or DPFP. For double-precision, this requires that $\mathbf{x}$ is converted to double when used, while the single-precision case is more straightforward. To characterise the difference, a similar test to Section~\ref{sec:comp} was used where points on the unit sphere, $\mathbb{S}^2$, were sampled and the mean error and maximum error were calculated. The error from points in $[-1,1]^3$ is also considered and the error is normalised by the vector magnitude before averaging and finding the maximum. Throughout these investigations we have used $\langle0,7,22\rangle\text{-}17\text{-}18$ compression. The results for a C++ implementation are shown in Table~\ref{tab:C++_error}. 

    \begin{figure}[tbhp]
		\centering
		\captionof{table}{Normalised error for $\mathbf{x}\in X$.}
		\begin{tabular}{c c c c c}	
			\toprule
			$\tan^{-1}$  & $\cos^{-1}$ & $\overline{e}$ & $\max{(e)}$ & $X$ \\
			\hline
 			Single & Single & $8.2832\times10^{-6}$ & $6.4805\times10^{-5}$ & $\mathbb{S}^2$\\ 
 			Single & Double & $8.2828\times10^{-6}$ & $1.7059\times10^{-5}$ & $\mathbb{S}^2$\\
 			Double & Single & $8.2831\times10^{-6}$ & $6.4805\times10^{-5}$ & $\mathbb{S}^2$\\ 
 			Double & Double & $8.2827\times10^{-6}$ & $1.7017\times10^{-5}$ & $\mathbb{S}^2$\\
 			\hline
 			Single & Single & $8.3016\times10^{-6}$ & $2.7380\times10^{-4}$ & $[-1,1]^3$\\
 			Single & Double & $8.3013\times10^{-6}$ & $1.7064\times10^{-5}$ & $[-1,1]^3$\\
 			Double & Single & $8.3014\times10^{-6}$ & $2.7380\times10^{-4}$ & $[-1,1]^3$\\
 			Double & Double & $8.3012\times10^{-6}$ & $1.7064\times10^{-5}$ & $[-1,1]^3$\\
 			\bottomrule
		\end{tabular}
		\label{tab:C++_error}
	\end{figure}
	
	Table~\ref{tab:C++_error} shows that the error is largely invariant to the intermediate precision of $\tan^{-1}$ used in the $\theta$ calculation. However, the error was found to be susceptible to the intermediate precision of $\phi$, where moving to single-precision caused a significant increase in the error. When calculating $\phi$ the quotient $z/r$ is required which serves as a potential source for catastrophic cancellation.  Any error in this quotient will then propagate through several subsequent steps in the compression routine. Two remedies were considered where DPFP was used for the division and the result was converted to SPFP before use in $\cos^{-1}$, as well as the division in SPFP and $\cos^{-1}$ in double. However, both gave the similar results as entirely SPFP processes. It is concluded that the calculation of $\phi$ should be performed in DPFP but $\theta$ can be performed in SPFP. To further highlight this, the value of $n_\theta$ and $n_\phi$ calculated in SPFP were compared to using DPFP. Table~\ref{tab:C++_misses} records the number of misses and it is clear that $\phi$ is significantly more lossy than $\theta$. 
	
    \begin{figure}[tbhp]
		\centering
		\captionof{table}{Single precision bin misses compared to double for $\mathbf{x}\in X$.}
		\begin{tabular}{r|c|l}	
			\toprule
			$n_\theta $ misses & $0.2008\%$ & $X = \mathbb{S}^2$ \\
			$n_\theta $ misses & $0.2039\%$ & $X = [-1,1]^3$ \\
			$n_\phi $ misses & $0.2411\%$ & $X = \mathbb{S}^2$ \\
			$n_\phi $ misses & $0.2603\%$ & $X = [-1,1]^3$ \\
 			\bottomrule
		\end{tabular}
		\label{tab:C++_misses}
	\end{figure}
	
	The effect of precision on the quantisation of $\theta$ and $\phi$ was also investigated. In the quantisation of $\phi$ it was found that the error and bin misses were invariant with precision. Therefore, after calculating $\phi$ in DPFP, it can be cast to a SPFP number for the quantisation. Similar results were found for $\theta$. A further, albeit small, reduction in the $\theta$ quantisation error was obtained by moving the domain for $\theta$ from $[0,2\pi]$ to $[-\pi,\pi]$. This avoids the addition, but the minus sign has to be detected and accounted for in the quantisation. 
	
	For completeness, these tests were also performed using CUDA-C, Fortran, and CUDA-Fortran using the same sample points. The results were found to be similar and in the case of the error terms were identical to the third and fourth significant figures. Sample implementations in the respective languages are included in the ESM.

\section{Applications to high-order CFD}\label{sec:numerics}

    The computational setting chosen to perform the numerical testing within is the flux reconstruction discontinuous spectral element method~\cite{Huynh2007,Huynh2009,Vincent2010}. This is a high-order method where the domain is meshed with elements within which polynomials are fitted. We will restrict the computational sub-domains here to be hexahedral, but other topologies are supported within FR. The internal solution point locations, as well as the interface flux point locations, will be a tensor product of Guass--Legendre quadrature points. The invisicid common interface flux is calculated using Rusanov's method with Davis wave speeds~\cite{Rusanov1961,Davis1988}. Furthermore, the common viscous interface flux is set using the BR1 method of Bassi and Rebay~\cite{Bassi1997a}. Throughout this section, temporal integration is performed by a 4th order low-storage explicit Runge--Kutta scheme~\cite{Kennedy2000}. The time step, $\Delta t$, was set such that the results were independent of the time scheme. 
    
    \begin{figure}[!p]
		\centering
		\captionof{table}{\label{tab:fr_nse}Steps for calculating flux divergence of Navier--Stokes with FR, with terms defined in accompanying text.  Quantities which may be subject to compression are highlighted in blue.}
		\begin{tabular}{r|rcl r}
		    \toprule
		    Step & Input & & Output & Description \\ \midrule
			1 & $\mathbf{q}^s$ & $\rightarrow$ & $\mathbf{q}^f$ & Interpolate solution to flux points\\
			2 & $[\mathbf{q}^f,$\mathhl{pale2}{\mathbf{n}_f}$]$ & $\rightarrow$ & $\mathbf{n}_f\cdot\mathbf{F}^I_\mathrm{inv}$ & Common interface invsicid flux\\
			3 & $\mathbf{q}^f$ & $\rightarrow$ & $\mathbf{w}^I$ & Common interface primitives\\
			4 & $\mathbf{q}^s$ & $\rightarrow$ & $\mathbf{w}^s$ & Primitives at solution points\\
			5 & $\mathbf{w}^s$ & $\rightarrow$ & \mathhl{pale2}{\nabla\mathbf{w}^s} & Gradient of primitives\\
			6 & $[$\mathhl{pale2}{\nabla\mathbf{w}^s}$,\mathbf{w}^I]$ & $\rightarrow$ & \mathhl{pale2}{\nabla\mathbf{w}^c} & Correct gradient of primitives\\
			7 & \mathhl{pale2}{\nabla\mathbf{w}^c} & $\rightarrow$ & \mathhl{pale2}{\nabla\mathbf{w}^f} & Interpolate solution to flux points\\
			8 & $[$\mathhl{pale2}{\nabla\mathbf{w}^f},\mathhl{pale2}{\mathbf{n}_f},\mathhl{pale2}{\mathbf{J}_f},$|\mathbf{J}_f|,\mathbf{q}^f]$ & $\rightarrow$ & $\mathbf{n}\cdot\mathbf{F}^I_\mathrm{vis}$ & Common interface viscous flux\\
			9 & $[$\mathhl{pale2}{\nabla\mathbf{w}^s},\mathhl{pale2}{\mathbf{J}_s}$,|\mathbf{J}_s|,\mathbf{q}^s]$ & $\rightarrow$ & \mathhl{pale2}{\mathbf{F}^s} & Flux at solution points\\
			10 & \mathhl{pale2}{\mathbf{F}^s} & $\rightarrow$ & $\nabla\cdot\mathbf{F}^s$ & Divergence of flux\\
			11 & $[$\mathhl{pale2}{\mathbf{F}^s}$,\mathbf{n}\cdot(\mathbf{F}^I_\mathrm{inv}-\mathbf{F}^I_\mathrm{vis})]$ & $ \rightarrow$ & $\mathbf{F}^{fc}$ & Correction to flux at interfaces\\
			12 & $[\mathbf{F}^{fc},\nabla\cdot\mathbf{F}^s]$ & $ \rightarrow$ & $\nabla\cdot\mathbf{F}^c$ & Correct divergence of flux\\
 		    \bottomrule
		\end{tabular}
	\end{figure}
    
    Table~\ref{tab:fr_nse} details the steps required to calculate the divergence of the flux (right-hand side) for the Navier--Stokes equations using the FR algorithm. Here $\mathbf{q}$ indicates the conserved variables, $\mathbf{w}$ the primitive variables, and $\mathbf{F}$ the flux tensor. Superscript $s$ indicates quantities at the solution points, $f$ at the flux points, $I$ the common interface values, $c$ corrected values at the solution points, and $cf$ the correction at the flux points.  Terms which can be subject to compression are highlighted in blue.
    
    Higher dimensional terms, such as fluxes and solution gradients, are handled by applying compression in a row-wise fashion.  For example, consider the inviscid flux tensor given by
    \begin{equation}\label{eq:eulerF}
			\mathbf{F} = \begin{tikzpicture}[baseline={([yshift=-.5ex]current bounding box.center)},vertex/.style={anchor=base,
    circle,fill=black!25,minimum size=18pt,inner sep=2pt}]
    			\matrix [matrix of math nodes,left delimiter={[},right delimiter={]}](m){
           			\rho u & \rho v & \rho w \\ 
           			\rho u^2 + p & \rho uv & \rho uw \\
					\rho uv & \rho v^2 + p & \rho vw \\
					\rho uw & \rho vw & \rho w^2 + p \\
					u(E+p) & v(E+p) & w(E+p) \\
        		};
        		\begin{scope}[on background layer]
        			\fill[{Pastel1-A},opacity=0.5] (m-5-1.west|-m-1-3.north) rectangle (m-1-3.south-|m-5-3.east);
        			\fill[{Pastel1-B},opacity=0.5] (m-5-1.west|-m-2-1.north) rectangle (m-2-3.south-|m-5-3.east);
        			\fill[{Pastel1-C},opacity=0.5] (m-5-1.west|-m-3-2.north) rectangle (m-3-3.south-|m-5-3.east);
        			\fill[{Pastel1-D},opacity=0.5] (m-5-1.west|-m-4-3.north) rectangle (m-4-3.south-|m-5-3.east);
        			\fill[{Pastel1-E},opacity=0.5] (m-5-1.west|-m-5-3.north) rectangle (m-5-3.south-|m-5-3.east);
        		\end{scope}
\end{tikzpicture},
	\end{equation}
	where $p$ is the pressure, $E$ is the total energy, and $\gamma$ is the ratio of specific heats.  Here, each colour-coded row is treated as a three component vector and compressed independently of any other row. 
	
	The means by which compression is incorporated into a kernel is shown in Alg.~\ref{alg:fr_fdiv_comp}.  This kernel, which corresponds to step 10 (divergence of the flux) applies the divergence operator, $\mathbf{D}$, to  the flux tensor, $\ob{F}^s = \mathrm{compress}(\mathbf{F}^s)$ at each nodal point, to give $\nabla\cdot\mathbf{F}^s$.   The decompression operation is highlighted in red.
    
    \begin{algorithm}[tbhp]
        \SetAlgoLined
        \KwData{$\ob{F}^S\in\mathbb{R}^{n_c\times n_s\times n_p}, \mathbf{D}\in\mathbb{R}^{3n_s\times n_s}$}
        \KwResult{$\nabla\cdot\mathbf{F}^s\in\mathbb{R}^{n_c\times n_s\times n_p}$}
        
        $\nabla\cdot\mathbf{F}^S \leftarrow 0$\\
        \tcp{Loop over elements}
        \For{$i \in [1,\dots, n_e]$}{
            \tcp{Outer loop over solution points}
            \For{$j \in [1,\dots, n_s]$}{
                \tcp{Decompress flux}
                \mathhl{pale1}{\mathbf{X} \leftarrow \mathrm{decompress}(\ob{F}^s_{ij})$ to give $\mathbf{X}\in\mathbb{R}^{n_p\times 3}}\\
                \tcp{Inner loop over solution points}
                \For{$k \in [1,\dots, n_s]$}{
                    \tcp{Prepare divergence operator}
                    $\mathbf{d} \leftarrow [\mathbf{D}_{jk},\mathbf{D}_{(j+n_s)k},\mathbf{D}_{(j+2n_s)k}]$\\
                    \tcp{Accumulate contribution to divergence}
                    $\nabla\cdot\mathbf{F}^s_{ik} \leftarrow  \nabla\cdot\mathbf{F}^s_{ik} + \mathbf{d}\cdot\mathbf{X}$
                }
            }
        }
        \caption{\label{alg:fr_fdiv_comp}Divergence of compressed flux (see Table~\ref{tab:fr_nse}, step 10). Difference from uncompressed kernel highlighted in red. Here $n_e$ is the number of elements, $n_s$ is the number of solution points, $n_f$ is the number of flux points, and $n_p=5$ is the number of equations.}
    \end{algorithm}
    
	
	\newpage
	For each step in the FR algorithm it is possible to directly calculate how much data must be read from and written out to memory.  The amount of data movement required by each step of the FR approach as described in Table~\ref{tab:fr_nse}, along with the reduction factor enabled by compression, can be seen in Table~\ref{tab:mem_usage}. These figures may be used to form a prediction for the performance improvement due compression an assumption which is well-supported by previous studies of FR on hexahedral domains~\cite{Witherden2014,Romero2020}. From this analysis we conclude that performance improvements of $20.1\%$, $20.6\%$, and $21.0\%$ for $k\in\{3,4,5\}$ respectively can be obtained. 
	
	\begin{figure}[tbhp]
		\centering
		\captionof{table}{\label{tab:mem_usage}Uncompressed memory transfer requirements ($T$ / KiB) per step per hexahedral element for FR applied to the Navier--Stokes Equations, and the memory compression ratio, $\Gamma$.}
		\begin{tabular}{r rc rc rc}
		    \toprule
		    \multirow{2}{0.8cm}{Step} & \multicolumn{2}{c}{$k=3$} & \multicolumn{2}{c}{$k=4$}  & \multicolumn{2}{c}{$k=5$}  \\ \cmidrule(lr){2-3} \cmidrule(lr){4-5} \cmidrule(l){6-7}
		    & \multicolumn{1}{c}{$T$} & $\Gamma$ & \multicolumn{1}{c}{$T$} & $\Gamma$ & \multicolumn{1}{c}{$T$} & $\Gamma$ \\ \midrule
			1 & \SI{3.13}{} & 1.00 & \SI{5.37}{} & 1.00 & \SI{8.44}{} & 1.00\\
			2 & \SI{4.88}{} & 0.92 & \SI{7.62}{} & 0.92 & \SI{10.97}{} & 0.92\\
			3 & \SI{3.75}{} & 1.00 & \SI{5.86}{} & 1.00 & \SI{8.44}{} & 1.00\\
			4 & \SI{2.50}{} & 1.00 & \SI{2.50}{} & 1.00 & \SI{2.50}{} & 1.00\\
			5 & \SI{5.00}{} & 0.75 & \SI{9.77}{} & 0.75 & \SI{16.88}{} & 0.75\\
			6 & \SI{9.38}{} & 0.73 & \SI{17.58}{} & 0.72 & \SI{29.53}{} & 0.71\\
			7 & \SI{9.38}{} & 0.67 & \SI{16.11}{} & 0.67 & \SI{25.31}{} & 0.67\\
			8 & \SI{16.13}{} & 0.79 & \SI{25.20}{} & 0.79 & \SI{36.28}{} & 0.79\\
			9 & \SI{11.25}{} & 0.71 & \SI{21.97}{} & 0.71 & \SI{37.97}{} & 0.71\\
			10 & \SI{5.00}{} & 0.75 & \SI{9.77}{} & 0.75 & \SI{16.88}{} & 0.75\\
			11 & \SI{7.50}{} & 0.83 & \SI{13.18}{} & 0.81 & \SI{21.09}{} & 0.80 \\
			12 & \SI{3.75}{} & 1.00 & \SI{7.32}{} & 1.00 & \SI{12.66}{} & 1.00\\ \midrule
			Weighted average & \SI{81.63}{} & 0.80 & \SI{142.25}{} & 0.79 & \SI{226.94}{} & 0.79 \\\bottomrule
		\end{tabular}
	\end{figure}
	
	\newpage
\subsection{Isentropic Convecting Vortex}
	
	We will begin our numerical experiments with the isentropic convecting vortex (ICV) as defined by Shu~\cite{Shu1997}. This is a solution to the compressible Euler's equations for an ideal gas, written  in 3D as
	\begin{subequations}
		\begin{align}
			\mathbf{q}_t + \nabla\cdot\mathbf{F} &= 0, \\
			\mathbf{q} &= \begin{bmatrix}
				\rho & \rho u & \rho v & \rho w & E
			\end{bmatrix}^T, \\
			E &= \half\rho(u^2 + v^2 + w^2) + \frac{p}{\gamma-1},
		\end{align}
	\end{subequations}
	where the flux tensor, $\mathbf{F}$, is defined in Eq.~(\ref{eq:eulerF}). The initial condition for the ICV was taken to be defined as
	\begin{subequations}
		\begin{align}
			u &= u_0 + \frac{\beta}{2\pi}(y_0-y)\exp{\bigg(\frac{1-r^2}{2}\bigg)},\\
			v &= v_0 - \frac{\beta}{2\pi}(x_0-x)\exp{\bigg(\frac{1-r^2}{2}\bigg)},\\
			w &= 0, \\
			p &= \bigg(1 - \frac{(\gamma-1)\beta^2}{8\gamma\pi^2}\exp{\big(1 - r^2\big)} \bigg)^\frac{\gamma}{\gamma-1},\\
			\rho &= p^{\frac{1}{\gamma}}, \\
			r &= \|\mathbf{x} - \mathbf{x}_0\|_2, 
		\end{align}
	\end{subequations}
	where $\beta$ is a constant controlling vortex strength, typically set to $\beta=5$. For the purposes of this test the convective velocities, $u_0$ and $v_0$, were set such that they vary on the unit circle, as
	\begin{equation}
		u_0 = \cos{\psi} \quad \text{and} \quad v_0 = \sin{\psi}.
	\end{equation}

	\begin{figure}[tbhp]
		\centering
			\resizebox{0.6\linewidth}{!}{\input{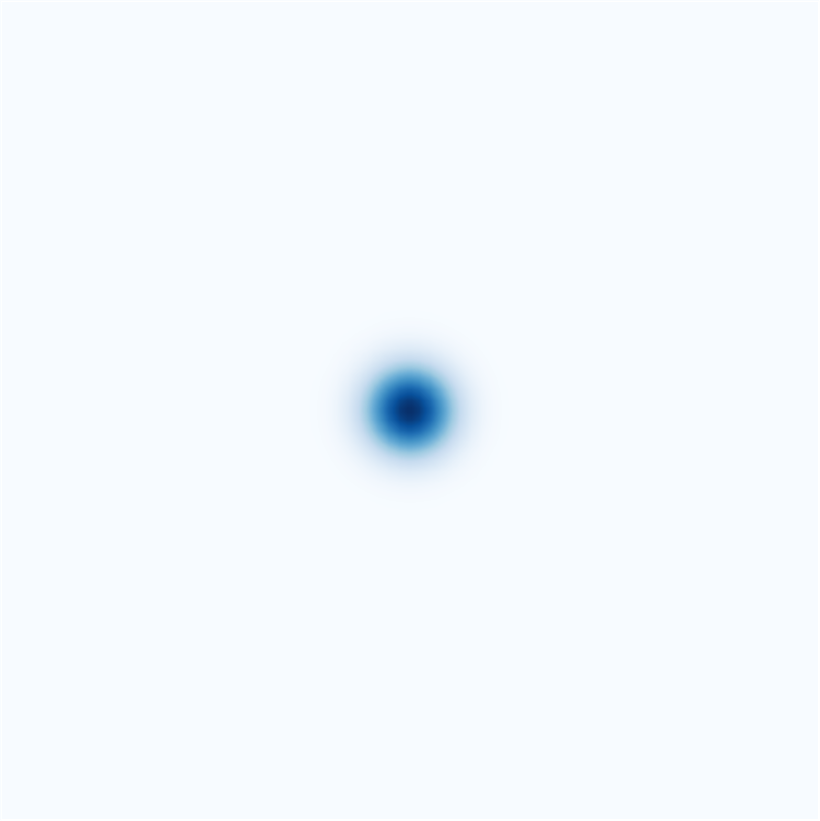}}
		\caption{ICV density showing orientation of convective velocity.}
		\label{fig:icv}
	\end{figure}
 	This flow was applied to a fully periodic domain with $\mathbf{\Omega} = [0,20]\times[0,20]\times[0,2]$, which was divided into $20\times20\times2$ subdomains.
	
\begin{figure}[tbhp]
		\centering
			\resizebox{0.7\linewidth}{!}{	\begin{tikzpicture}
		\begin{axis}[name=plot1,xlabel={$\psi$},ylabel={$e_1$},
		    xtick={0,30,60,90},ytick={0.5e-4,1e-4,1.5e-4,2e-4},
		    xticklabels={$0$,$30$,$60$,$90$},
    		xmin=0,xmax=90,
		    ylabel style={rotate=-90},
    		y tick label style={
        		/pgf/number format/.cd,
            	fixed,
            	fixed zerofill,
            	precision=1,
        	/tikz/.cd
    		},
    		legend style={at={(0.5,0.97)},anchor=north,font=\scriptsize},
    		ymin=0.3e-4,ymax=2e-4,
    		style={font=\large}]			
			\addplot[color={Set1-E}, style={thick}]
				table[x=theta,y=eb,col sep=comma,unbounded coords=jump]{./Figs/data/icv_error_comp.csv};
			\addlegendentry{Base}
			
			\addplot[color={Set1-C}, style={thick}]
				table[x=theta,y=e1616,col sep=comma,unbounded coords=jump]{./Figs/data/icv_error_comp.csv};
			\addlegendentry{$\langle1,8,23\rangle\mbox{-}16\mbox{-}16$}
			
			\addplot[color={Set1-A}, style={thick}]
				table[x=theta,y=e1617,col sep=comma,unbounded coords=jump]{./Figs/data/icv_error_comp.csv};
			\addlegendentry{$\langle0,8,23\rangle\text{-}16\text{-}17$}
			
			\addplot[color={Set1-B}, style={thick}]
				table[x=theta,y=e1717,col sep=comma,unbounded coords=jump]{./Figs/data/icv_error_comp.csv};
			\addlegendentry{$\langle0,7,23\rangle\mbox{-}17\mbox{-}17$}
			
			\addplot[color={Set1-D}, style={thick}]
				table[x=theta,y=e1718,col sep=comma,unbounded coords=jump]{./Figs/data/icv_error_comp.csv};
			\addlegendentry{$\langle0,7,22\rangle\mbox{-}17\mbox{-}18$}
		\end{axis} 		
	\end{tikzpicture}}
		\caption{ICV error with flow angle for various methods for degree $k=4$ FR.}
		\label{fig:icv_error}
	\end{figure}
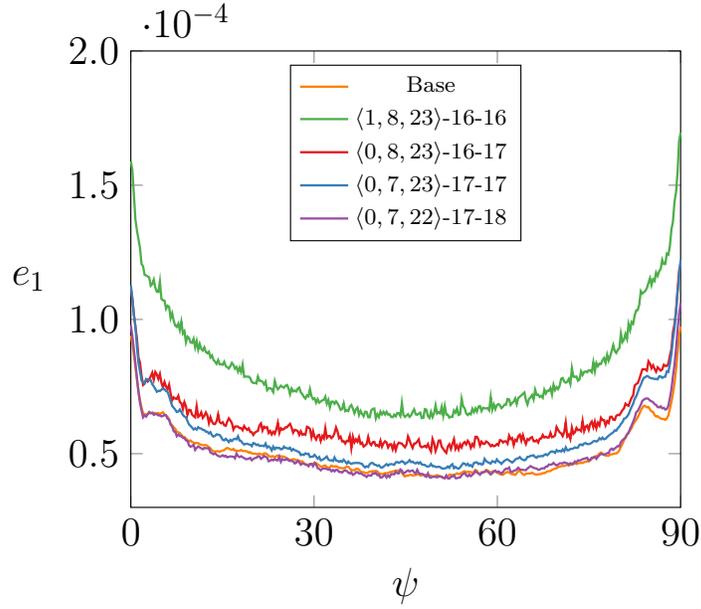

	The ICV test case permits the actual solution error to be analytically calculated at later times, and this is the primary advantage of using this test case. Here the results presented will focus on the absolute error in $\rho$ averaged over the whole domain, denoted as $e_1$ and defined as
	\begin{equation}
		e_1(t) = \frac{1}{N}\sum_{i=1}^{N}\|\rho_\mathrm{exact}(t,\mathbf{x}_i) - \rho(t,\mathbf{x}_i)\|_1
	\end{equation}		
	where $N$ is the number of points used for averaging. Here we have averaged over all the solution points in the domain, which for the grid defined is $20\times20\times2\times(4+1)^3=10^5$. The error was considered at $t=20$, which for $\psi=\{0,90\}$ should lead to the vortex returning to the start position. The error as the convective velocity angle is varied is displayed in Fig.~\ref{fig:icv_error}; the line labelled as `base' is the behaviour for the unaltered scheme running in single precision. Anisotropy of the maximal-order polynomial basis used in this tensor product formulation is clearly visible as a reduction in the error by approximately half as the angle is varied. 
	
	Going on to examine the impact of compression, it is clear the additional bits obtained from the sign, exponent, and mantissa have a significant effect on the error of the scheme, with $\langle0,7,22\rangle\text{-}17\text{-}18$ having comparative error to the base scheme.

\subsection{Taylor--Green Vortex}
	
	We will now move on to consideration of the more complex compressible Navier--Stokes Equation set, written in the form
	\begin{subequations}
		\begin{align}
			&\mathbf{q}_t + \nabla\cdot\mathbf{F} = \nabla\cdot\mathbf{F}^v, \quad \mathrm{where}\\
			&\mathbf{F}^v = \begin{bmatrix}
				0 & 0 & 0 \\
				\tau_{xx} & \tau_{yx} & \tau_{zx} \\
				\tau_{xy} & \tau_{yy} & \tau_{zy} \\
				\tau_{xz} & \tau_{yz} & \tau_{zz} \\
				u\tau_{xx} + v\tau_{xy} + w\tau_{xz} - b_x & u\tau_{yx} + v\tau_{yy} + w\tau_{yz} - b_y & u\tau_{zx} + v\tau_{zy} + w\tau_{zz} - b_z
			\end{bmatrix}
		\end{align}
	\end{subequations}
	Here $\tau_{ij}$ is the standard viscous stress tensor and $b_i$ are the components of $-\kappa\nabla T$, where $T$ is the temperature. Setting the bulk viscosity to zero, these may be written as
	\begin{subequations}
		\begin{align}
		\tau_{xy} &= \tau_{yx} = \mu(u_y + v_x), \quad \tau_{xx} = 2\lambda u_x - \lambda(v_y + w_z),\\
		\tau_{yz} &= \tau_{zy} = \mu(v_z + w_y), \quad \tau_{yy} = 2\lambda v_y - \lambda(w_z + u_x),\\
		\tau_{zx} &= \tau_{xz} = \mu(w_x + u_z), \quad \tau_{zz} = 2\lambda w_z - \lambda(u_x + v_y),\\
		\mathbf{b} &= [b_x,b_y,b_z]^T = -\kappa\nabla T,
		\end{align}
	\end{subequations}	
	where $\lambda=2\mu/3$ and $\kappa$ is the thermal diffusivity.	This equation set, after application to the FR methodology, offers several opportunities for compression. Again the flux terms can be compressed, as in Eq.~(\ref{eq:eulerF}), with compression being applied to the aggregate of $\mathbf{F}-\mathbf{F}^v$ with compression applied for these flux terms at both the FR solution and flux points. Furthermore, the gradient terms used in the construction of $\tau_{ij}$ can be compressed once again at the FR solution and flux points.
	
	The case which was applied to the Navier--Stokes equations was the Taylor--Green vortex~\cite{Taylor1937,Brachet1983}, whose initial condition is defined with
	\begin{equation}\label{eq:tgv_xyz}
		\mathbf{u} = \begin{bmatrix}
			U_0f(x,y,z) \\
			-U_0f(y,x,z) \\
			0
		\end{bmatrix}, \quad p = p(x,y,z).	
	\end{equation}
	Here we have defined the functions $f$ and $p$ for simplicity, and they take the form: 
	\begin{subequations}
		\begin{align}
			f(x,y,z) &= \sin{\frac{x}{k}}\cos{\frac{y}{k}}\cos{\frac{z}{k}}, \\
			p(x,y,z) &= p_0 + \frac{\rho_0U_0^2}{16}\bigg(\cos{\frac{2x}{k}} + \cos{\frac{2y}{k}}\bigg)\bigg(2 + \cos{\frac{2z}{k}}\bigg).
		\end{align}
	\end{subequations}
	The density is then set based on Boyle's Law, assuming an ideal gas,
	\[
		\rho = \frac{p\rho_0}{p_0}.
	\]
	We introduced these functions in the definition of the TGV as it permits clear notation for spatial transformation of the initial condition with the condition in Eq.~(\ref{eq:tgv_xyz}) being traditionally used. The proposed compression was shown in Section~\ref{sec:comp} to have some degree on anisotropy. Hence, we also investigated a slight modification of the initial condition
	\begin{equation}\label{eq:tgv_zyx}
		\mathbf{u} = \begin{bmatrix}
			U_0f(y,x,z) \\
			-U_0f(x,y,z) \\
			0
		\end{bmatrix}, \quad p = p(z,y,x).			
	\end{equation}
	This condition exhibits a rotation through $90^\circ$. Evaluation was performed using the following dissipation metrics:
	\begin{align}
		\xi_1 &= -\frac{1}{\rho_0U_0^2|\mathbf{\Omega}|}\dx{}{t}\int_\mathbf{\Omega}\half\rho(\mathbf{x},t)(\mathbf{u}\cdot\mathbf{u})\mathrm{d}\mathbf{x}, \\
		\xi_2 &= \frac{1}{2\mu U_0^2|\mathbf{\Omega}|}\int_\mathbf{\Omega}\half\rho(\mathbf{x},t)(\pmb{\omega}\cdot\pmb{\omega})\mathrm{d}\mathbf{x}, \quad \mathrm{where} \quad \pmb{\omega} = \nabla\times\mathbf{u}.
	\end{align}
	Throughout, a TGV was considered for $\mathbf{\Omega} = [0,2\pi]^3$ with $Re=1600$, $Ma=0.08$ and $Pr=0.71$, based on $U_0=1$, $k=1$, $\rho_0=1$ and $p_0=100$. Furthermore, low-storage explicit four-stage fourth-order Runge--Kutta time integration was used and a time step was chosen such that results were independent of $\Delta t$. For the cases investigated here, that meant a value of $\Delta t = 10^{-3}$. Comparative DNS results were obtained from DeBonis~\cite{DeBonis2013}.
	
	\begin{figure}[tbhp]
		\centering
		\begin{subfigure}[h]{0.8\linewidth}
			\centering
				\begin{tikzpicture}
		\begin{axis}[name=plot1,xlabel={$t$},ylabel={$\xi_2$},
		    xtick={0,5,...,20},ytick={0,0.002,0.004,0.006,0.008,0.01,0.012,0.014,0.016},
    		xmin=-0,xmax=20,
    		ymin=0,ymax=0.015,
		    ylabel style={rotate=-90},
    		y tick label style={
        		/pgf/number format/.cd,
            	fixed,
            	fixed zerofill,
            	precision=1,
        	/tikz/.cd
    		},
    		legend style={at={(0.97,0.97)},anchor=north east,font=\scriptsize}]
			\addplot[black, dotted, style={thick}]
				table[x index={0},y index={1},col sep=comma,unbounded coords=jump]{./Figs/data/TGV_ref_spec_1600.csv};
			\addlegendentry{DNS}
			
			\addplot[color={Set1-A}, style={thick}]
				table[x=t,y=base,col sep=comma,unbounded coords=jump]{./Figs/data/tgv_comp_ens.csv};
			\addlegendentry{Base}

			\addplot[color={Set1-B}, style={thick}]
				table[x=t,y=1716,col sep=comma,unbounded coords=jump]{./Figs/data/tgv_comp_ens.csv};
			\addlegendentry{$\langle0,8,23\rangle\text{-}16\text{-}17$}

			\addplot[color={Set1-C}, style={thick}]
				table[x=t,y=1717,col sep=comma,unbounded coords=jump]{./Figs/data/tgv_comp_ens.csv};
			\addlegendentry{$\langle0,7,23\rangle\mbox{-}17\mbox{-}17$}

			\addplot[color={Set1-D}, style={thick}]
				table[x=t,y=1817,col sep=comma,unbounded coords=jump]{./Figs/data/tgv_comp_ens.csv};
			\addlegendentry{$\langle0,7,22\rangle\mbox{-}17\mbox{-}18$}
		\end{axis} 		
	\end{tikzpicture}
			\caption{$\mathbf{u} = U_0[f(x,y,z),-f(y,x,z),0]^T$}
			\label{fig:TGVxyz_global}
		\end{subfigure}
		~
		\begin{subfigure}[h]{0.8\linewidth}
			\centering
				\begin{tikzpicture}
		\begin{axis}[name=plot1,xlabel={$t$},ylabel={$\xi_2$},
		    xtick={7,8,...,12},ytick={0.008,0.009,0.01,0.011},
    		xmin=7,xmax=12,
    		ymin=0.0075,ymax=0.0111,
		    ylabel style={rotate=-90},
    		y tick label style={
        		/pgf/number format/.cd,
            	fixed,
            	fixed zerofill,
            	precision=1,
        	/tikz/.cd
    		},
    		legend style={at={(0.45,0.05)},anchor=south,font=\scriptsize}]
			\addplot[black, dotted, style={thick}]
				table[x index={0},y index={1},col sep=comma,unbounded coords=jump]{./Figs/data/TGV_ref_spec_1600.csv};
			\addlegendentry{DNS}
			
			\addplot[color={Set1-A}, style={thick}]
				table[x=t,y=base,col sep=comma,unbounded coords=jump]{./Figs/data/tgv_comp_ens.csv};
			\addlegendentry{Base}

			\addplot[color={Set1-B}, style={thick}]
				table[x=t,y=1716,col sep=comma,unbounded coords=jump]{./Figs/data/tgv_comp_ens.csv};
			\addlegendentry{$\langle0,8,23\rangle\text{-}16\text{-}17$}

			\addplot[color={Set1-C}, style={thick}]
				table[x=t,y=1717,col sep=comma,unbounded coords=jump]{./Figs/data/tgv_comp_ens.csv};
			\addlegendentry{$\langle0,7,23\rangle\mbox{-}17\mbox{-}17$}

			\addplot[color={Set1-D}, style={thick}]
				table[x=t,y=1817,col sep=comma,unbounded coords=jump]{./Figs/data/tgv_comp_ens.csv};
			\addlegendentry{$\langle0,7,22\rangle\mbox{-}17\mbox{-}18$}
		\end{axis} 		
	\end{tikzpicture}
			\caption{$\mathbf{u} = U_0[f(x,y,z),-f(y,x,z),0]^T$}
			\label{fig:TGVxyz}
		\end{subfigure}
		\caption{Taylor--Green Vortex enstrophy based decay rate, compression comparison for $Re=1600$ and $Ma=0.08$ on a $16^3$ element hexahedral grid with degree $k=4$ FR.}
		\label{fig:TGV_comp}
	\end{figure}
	
	Initially, a comparison is made between the three compression regimes using the initial condition set out in Eq.~(\ref{eq:tgv_xyz}).  The results can be seen in Fig.~\ref{fig:TGV_comp}. This particular flow is initially laminar and transition to turbulence occurs around $3<t<7$.  Looking at the figure, it is evident that compression has had a limited impact on the laminar region, which may have been predicted from the ICV results. Throughout transition as turbulent structures develop small scale motions begin to have an increased importance. In particular, the compression method will introduce small perturbations, which can affect the energy at the smallest scales. It can be seen from Fig.~\ref{fig:TGVxyz} that compression has led to a small reduction in the peak turbulent dissipation. This may be attributed to the compression error having the greatest impact at the smallest scales, the scales are known to have a large contribution to the scheme dissipation, as was demonstrated by Trojak~\etal~\cite{Trojak2018d}. The TGV results here conform with ICV tests with the $\langle0,7,22\rangle\text{-}17\text{-}18$ compression regime introducing the least error.
	
	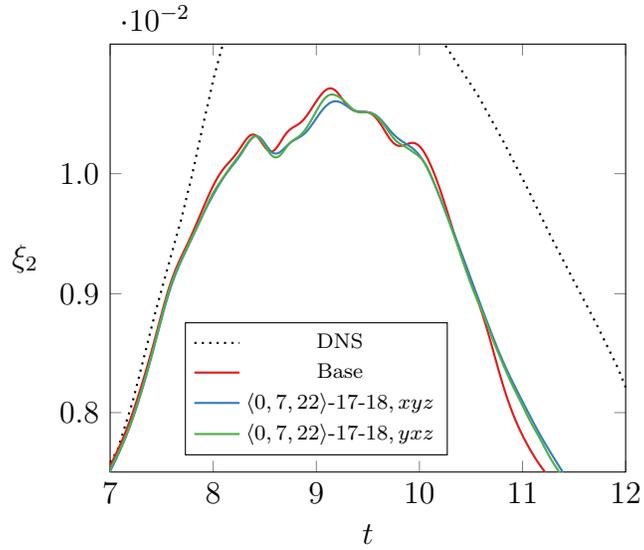
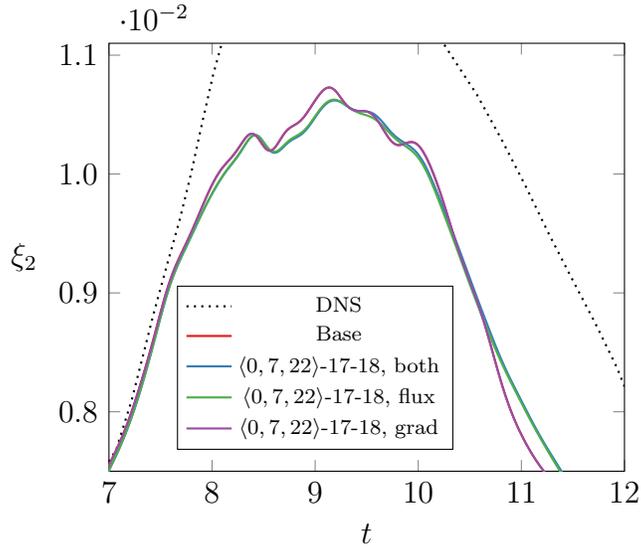
\begin{figure}[tbhp]
		\centering
		\begin{subfigure}[h]{0.8\linewidth}
			\centering
				\begin{tikzpicture}
		\begin{axis}[name=plot1,xlabel={$t$},ylabel={$\xi_2$},
		    xtick={7,8,...,12},ytick={0.008,0.009,...,0.011},
    		xmin=7,xmax=12,
    		ymin=0.0075,ymax=0.0111,
		    ylabel style={rotate=-90},
    		y tick label style={
        		/pgf/number format/.cd,
            	fixed,
            	fixed zerofill,
            	precision=1,
        	/tikz/.cd
    		},
    		legend style={at={(0.4,0.35)},anchor=north,font=\scriptsize}]
			\addplot[black, dotted, style={thick}]
				table[x index={0},y index={1},col sep=comma,unbounded coords=jump]{./Figs/data/TGV_ref_spec_1600.csv};
			\addlegendentry{DNS}
			
            \addplot[color={Set1-A}, style={thick}]
				table[x=t,y index={1},col sep=comma,unbounded coords=jump]{./Figs/data/tgv_comp_ens.csv};
			\addlegendentry{Base}

			\addplot[color={Set1-B}, style={thick}]
				table[x=t,y index={4},col sep=comma,unbounded coords=jump]{./Figs/data/tgv_comp_ens.csv};
			\addlegendentry{$\langle0,7,22\rangle\text{-}17\text{-}18, xyz$}

            \addplot[color={Set1-C}, style={thick}]
				table[x=t,y index={5},col sep=comma,unbounded coords=jump]{./Figs/data/tgv_comp_ens.csv};
			\addlegendentry{$\langle0,7,22\rangle\mbox{-}17\mbox{-}18, yxz$}
		\end{axis} 		
	\end{tikzpicture}
	
			\caption{Comparison of Eq.~(\ref{eq:tgv_xyz}) and Eq.~(\ref{eq:tgv_zyx}).}
			\label{fig:TGV_xyz_xzy}
		\end{subfigure}
		~
		\begin{subfigure}[h]{0.8\linewidth}
			\centering
				\begin{tikzpicture}
		\begin{axis}[name=plot1,xlabel={$t$},ylabel={$\xi_2$},
		    xtick={7,8,...,12},ytick={0.008,0.009,0.010,0.011},
    		xmin=7,xmax=12,
    		ymin=0.0075,ymax=0.0111,
		    ylabel style={rotate=-90},
    		y tick label style={
        		/pgf/number format/.cd,
            	fixed,
            	fixed zerofill,
            	precision=1,
        	/tikz/.cd
    		},
    		legend style={at={(0.4,0.05)},anchor=south,font=\scriptsize}]
			\addplot[black, dotted, style={thick}]
				table[x index={0},y index={1},col sep=comma,unbounded coords=jump]{./Figs/data/TGV_ref_spec_1600.csv};
			\addlegendentry{DNS};
			
			\addplot[color={Set1-A}, style={thick}]
				table[x=t,y index={1},col sep=comma,unbounded coords=jump]{./Figs/data/tgv_comp_ens.csv};
			\addlegendentry{Base};

			\addplot[color={Set1-B}, style={thick}]
				table[x=t,y index={4},col sep=comma,unbounded coords=jump]{./Figs/data/tgv_comp_ens.csv};
			\addlegendentry{$\langle0,7,22\rangle\text{-}17\text{-}18$, both};

			\addplot[color={Set1-C}, style={thick}]
				table[x=t,y index={7},col sep=comma,unbounded coords=jump]{./Figs/data/tgv_comp_ens.csv};
			\addlegendentry{$\langle0,7,22\rangle\mbox{-}17\mbox{-}18$, flux};

			\addplot[color={Set1-D}, style={thick}]
				table[x=t,y index={6},col sep=comma,unbounded coords=jump]{./Figs/data/tgv_comp_ens.csv};
			\addlegendentry{$\langle0,7,22\rangle\mbox{-}17\mbox{-}18$, grad};
			
		\end{axis} 		
	\end{tikzpicture}
			\caption{Flux vs. gradient compression}
			\label{fig:TGV_fr}
		\end{subfigure}
		\caption{Taylor--Green Vortex enstrophy based decay rate, transformed and partial compression comparison for $Re=1600$ and $Ma=0.08$ on a $16^3$ element hexahedral grid with degree $k=4$ FR.}
		\label{fig:TGV_comp2}
	\end{figure}
	
	As was noted in Section~\ref{ssec:bits}, the proposed compression method has some degree of anisotropy. To understand if this anisotropy has a significant impact on the TGV test case the initial condition was transformed to that of Eq.~(\ref{eq:tgv_zyx}). Comparison is displayed in Fig.~\ref{fig:TGV_xyz_xzy} for the compression regime $\langle0,7,22\rangle\text{-}17\text{-}18$. It is clear that anisotropy has affected the results in this case, but the difference is small compared to the potential difference caused by reduction to $\langle0,8,23\rangle\text{-}16\text{-}17$. Therefore, the degree of isotropy achieved by $\langle0,7,22\rangle\text{-}17\text{-}18$ is deemed to be acceptable.
	
	Within FR there is the potential for compression to be applied to either the flux terms, gradient terms, or both of these components. As gradient terms are solely used in the calculation of the viscous flux, and as the viscous flux typically has significantly smaller magnitude than the inviscid flux, it is hypothesised that compressing the flux will have a larger impact than the gradient in this case. Figure~\ref{fig:TGV_fr} makes this comparison. It may be observed that the differences are of the order of the differences observed in Fig.~\ref{fig:TGV_comp} and the application of compression of both terms leads to smaller observed differences.

    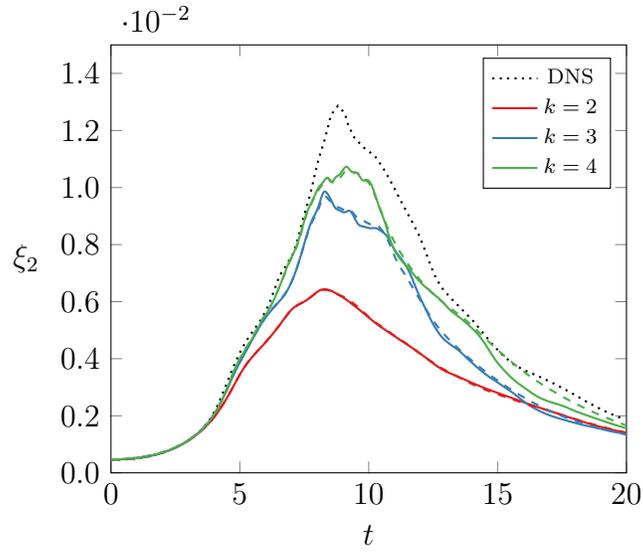
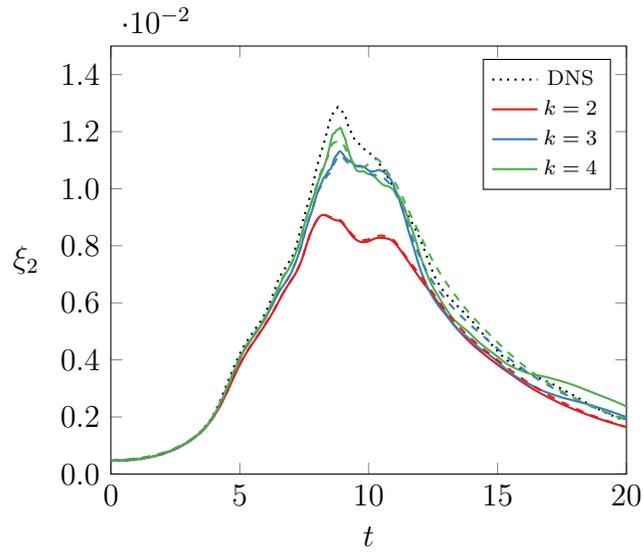
\begin{figure}[tbhp]
		\centering
		\begin{subfigure}[h]{0.8\linewidth}
			\centering
				\begin{tikzpicture}
		\begin{axis}[name=plot1,xlabel={$t$},ylabel={$\xi_2$},
		    xtick={0,5,...,20},ytick={0,0.002,0.004,0.006,0.008,0.01,0.012,0.014,0.016},
    		xmin=-0,xmax=20,
    		ymin=0,ymax=0.015,
		    ylabel style={rotate=-90},
    		y tick label style={
        		/pgf/number format/.cd,
            	fixed,
            	fixed zerofill,
            	precision=1,
        	/tikz/.cd
    		},
    		legend style={at={(0.97,0.97)},anchor=north east,font=\scriptsize}]
			\addplot[black, dotted, style={thick}]
				table[x index={0},y index={1},col sep=comma,unbounded coords=jump]{./Figs/data/TGV_ref_spec_1600.csv};
			\addlegendentry{DNS};
			
			\addplot[color={Set1-A}, style={thick}]
				table[x=t,y=p2x27,col sep=comma,unbounded coords=jump]{./Figs/data/tgv_lowres_ens.csv};
			\addlegendentry{$k=2$};
			
			\addplot[color={Set1-B}, style={thick}]
				table[x=t,y=p3x20,col sep=comma,unbounded coords=jump]{./Figs/data/tgv_lowres_ens.csv};
			\addlegendentry{$k=3$};
			
			\addplot[color={Set1-C}, style={thick}]
				table[x=t,y=p4x16,col sep=comma,unbounded coords=jump]{./Figs/data/tgv_lowres_ens.csv};
			\addlegendentry{$k=4$};
			
			\addplot[color={Set1-A}, style={thick,dashed}]
				table[x=t,y=p2x27c,col sep=comma,unbounded coords=jump]{./Figs/data/tgv_lowres_ens.csv};
			
			\addplot[color={Set1-B}, style={thick,dashed}]
				table[x=t,y=p3x20c,col sep=comma,unbounded coords=jump]{./Figs/data/tgv_lowres_ens.csv};
			
			\addplot[color={Set1-C}, style={thick,dashed}]
				table[x=t,y=p4x16c,col sep=comma,unbounded coords=jump]{./Figs/data/tgv_lowres_ens.csv};
		\end{axis} 		
	\end{tikzpicture}
			\caption{\label{fig:tgv_lowres}$\mathrm{DoF}\approx80^3$.}
		\end{subfigure}
		~
		\begin{subfigure}[h]{0.8\linewidth}
			\centering
			    \begin{tikzpicture}
		\begin{axis}[name=plot1,xlabel={$t$},ylabel={$\xi_2$},
		    xtick={0,5,...,20},ytick={0,0.002,0.004,0.006,0.008,0.01,0.012,0.014,0.016},
    		xmin=-0,xmax=20,
    		ymin=0,ymax=0.015,
		    ylabel style={rotate=-90},
    		y tick label style={
        		/pgf/number format/.cd,
            	fixed,
            	fixed zerofill,
            	precision=1,
        	/tikz/.cd
    		},
    		legend style={at={(0.97,0.97)},anchor=north east,font=\scriptsize}]
			\addplot[black, dotted, style={thick}]
				table[x index={0},y index={1},col sep=comma,unbounded coords=jump]{./Figs/data/TGV_ref_spec_1600.csv};
			\addlegendentry{DNS}
			
			\addplot[color={Set1-A}, style={thick}]
				table[x=t,y=p2x53,col sep=comma,unbounded coords=jump]{./Figs/data/tgv_hires_ens.csv};
			\addlegendentry{$k=2$};
			
			\addplot[color={Set1-B}, style={thick}]
				table[x=t,y=p3x40,col sep=comma,unbounded coords=jump]{./Figs/data/tgv_hires_ens.csv};
			\addlegendentry{$k=3$};
			
			\addplot[color={Set1-C}, style={thick}]
				table[x=t,y=p4x32,col sep=comma,unbounded coords=jump]{./Figs/data/tgv_hires_ens.csv};
			\addlegendentry{$k=4$};
			
			\addplot[color={Set1-A}, style={thick,dashed}]
				table[x=t,y=p2x53c,col sep=comma,unbounded coords=jump]{./Figs/data/tgv_hires_ens.csv};
			
			\addplot[color={Set1-B}, style={thick,dashed}]
				table[x=t,y=p3x40c,col sep=comma,unbounded coords=jump]{./Figs/data/tgv_hires_ens.csv};
			
			\addplot[color={Set1-C}, style={thick,dashed}]
				table[x=t,y=p4x32c,col sep=comma,unbounded coords=jump]{./Figs/data/tgv_hires_ens.csv};
		\end{axis} 		
	\end{tikzpicture}
			\caption{\label{fig:tgv_hires}$\mathrm{DoF}\approx160^3$.}
		\end{subfigure}
		\caption{\label{fig:TGV_res}Taylor--Green Vortex enstrophy derived dissipation for various $k$ and DoF. Solid lines indicate the uncompressed version, and dashed the compressed with $\langle0,7,22\rangle\text{-}17\text{-}18$.}
	\end{figure}
	Finally, to demonstrate the effect of compression with varying $k$ and $h$, tests were conducted for $k\in\{2,3,4\}$ and $\mathrm{DoF}\in\{80^3,160^3\}$. The outcomes of which are presented in Fig.~\ref{fig:TGV_res} where small variations in the enstophy between the compressed and uncompressed scheme can be observed. When considering the highest resolution case of $k=4$ and $\mathrm{DoF}=160^3$, the increased fidelity made the difference more pronounced. However, we remark here that the differences are similar in magnitude to what one would observe when changing the approximate Riemann solver.
\section{Performance Evaluation}\label{sec:performance}
    In this section we will evaluate the computational performance of our proposed scheme. For this, the following problem was considered:
	\begin{equation}
		\mathbf{C} = \mathbf{A} + \mathbf{B},
	\end{equation}
	where $\mathbf{A}$, $\mathbf{B}$, and $\mathbf{C}$ are $3 \times N$ matrices of single-precision numbers. This case was chosen as it has an extremely low arithmetic intensity.
	
	A kernel was designed that took in two 64-bit unsigned integers and returned one 64-bit unsigned integer. Within the kernel, the inputs were decompressed into two 3D vectors, the vectors were added, and the result was compressed to give the output. Throughout $\langle0,7,22\rangle\text{-}17\text{-}18$ compression was used.   All benchmarks were performed on an NVIDIA Titan V GPU which is representative of the hardware typically employed by high performance FR codes.  The specifications of the GPU can be seen in Table.~\ref{tab:titanv_spec}.
	
	\begin{figure}[tbhp]
		\centering
		\captionof{table}{\label{tab:titanv_spec}NVIDIA Titan V key statistics.}
		\begin{tabular}{r|l}	
			\toprule
			Global Memory & \SI{12.6}{\giga\byte}\\
 			L2 Cache & \SI{4.5}{\mebi\byte} \\ 
 			Clock Rate & \SI{1.455}{\giga\hertz} \\ 
 			Registers/Block & 65536 \\ 
 			 \bottomrule
		\end{tabular}
	\end{figure}	
	
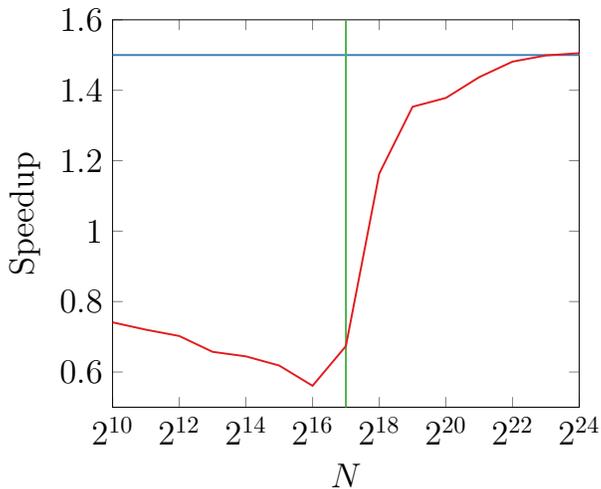
\begin{figure}[tbhp]
		\centering
		\resizebox{0.6\linewidth}{!}{	\begin{tikzpicture}
		\begin{axis}[name=plot1,xlabel={$N$},ylabel={Speedup},
		    xtick={10,12,...,24}, 
		    xticklabels={$2^{10}$,$2^{12}$,$2^{14}$,$2^{16}$,$2^{18}$,$2^{20}$,$2^{22}$,$2^{24}$},
    		xmin=10,xmax=24,
    		ymin=0.5,ymax=1.6,
    		style={font=\large}]
			\addplot[color={Set1-B}, style={thick}] coordinates {(10,1.5) (24,1.5)};
			\addplot[color={Set1-C}, style={thick}] coordinates {(17,0.5) (17,1.6)};
			\addplot[color={Set1-A}, style={thick}]
				table[x index={0},y index={1},unbounded coords=jump]{./Figs/data/runtime_optt.dat};
		\end{axis} 		
	\end{tikzpicture}}
		\caption{Speedup of compression vs. array size on NVIDIA Titan V GPU shown in red. The additional blue line corresponds to theoretical maximum speedup of 1.5 whilst the the array size corresponding to the 4.5MiB L2 cache is shown in green.}
		\label{fig:speedup}
	\end{figure}

	The results comparing the speedup afforded by the kernel incorporating compression as matrix size varies are shown in Fig.~\ref{fig:speedup}. Included are lines to indicate the theoretical peak speedup (dashed line), as well as a line indicating the baseline problem size which exceeds the capacity of the L2 cache (dotted line). It is clear that once the matrices may no longer be entirely resident in L2 cache, the compression kernel becomes more favourable. This proves our hypothesis that this form of compression may aid in reducing computation time in applications that are bound by global memory bandwidth. Furthermore, by $N=2^{22}$ close to peak speedup is obtained, which is equivalent to \SI{144}{\mebi\byte} of data in the baseline case. We remark here that \SI{144}{\mebi\byte} is a relatively small amount of data relative to the size of global memory.  As such, we expect a large proportion of bandwidth bound kernels to be able to achieve the optimal speedup.
	
\section{Conclusions}
\label{sec:conclusions}
	A compression algorithm was presented for three-dimensional single-precision vectors that was designed to work on local data to reduce memory bandwidth when retrieving vectors from global memory. The method utilised spherical polar coordinates, discretising the angles using integers, and using a modified floating-point bit layout. It was found that a seven-bit exponent and 22-bit mantissa were sufficient to increase the quantisation of the angles to such a point that in numerical test cases on Euler's equations, the error was appreciably similar to the baseline scheme. Further tests were performed on the compressible Navier--Stokes equations, particularly the transitional Taylor-Green test case. Only limited variation in results was observed when compression was applied in this case. Lastly, performance evaluation showed that a speedup close to the theoretical peak could be achieved for cases that could not be entirely resident in the L2 cache.

\section*{Acknowledgements}
The authors would like to thank Chris Cox, Tarik Dzanic, and Lai Wang for their technical  editing,  language  editing,  and proofreading.
\label{sec:ack}

\section*{References}
\bibliographystyle{elsarticle-num}
\bibliography{library}

\begin{thebibliography}{10}
\expandafter\ifx\csname url\endcsname\relax
  \def\url#1{\texttt{#1}}\fi
\expandafter\ifx\csname urlprefix\endcsname\relax\def\urlprefix{URL }\fi
\expandafter\ifx\csname href\endcsname\relax
  \def\href#1#2{#2} \def\path#1{#1}\fi

\bibitem{Witherden2014}
F.~D. Witherden, A.~M. Farrington, P.~E. Vincent,
  \href{http://dx.doi.org/10.1016/j.cpc.2014.07.011}{{PyFR: An Open Source
  Framework for Solving Advection-Diffusion Type Problems on Streaming
  Architectures Using the Flux Reconstruction Approach}}, Computer Physics
  Communications 185~(11) (2014) 3028--3040.
\newblock \href {http://arxiv.org/abs/1312.1638} {\path{arXiv:1312.1638}},
  \href {http://dx.doi.org/10.1016/j.cpc.2014.07.011}
  {\path{doi:10.1016/j.cpc.2014.07.011}}.
\newline\urlprefix\url{http://dx.doi.org/10.1016/j.cpc.2014.07.011}

\bibitem{Witherden2019}
F.~D. Witherden, A.~Jameson, {Impact of Number Representation for High-Order
  Implicit Large-Eddy Simulations}, AIAA Journal 58~(1) (2019) 1--14.
\newblock \href {http://dx.doi.org/10.2514/1.j058434}
  {\path{doi:10.2514/1.j058434}}.

\bibitem{Homann2007}
H.~Homann, J.~Dreher, R.~Grauer, {Impact of the Floating-Point Precision and
  Interpolation Scheme on the Results of DNS of Turbulence by Pseudo-Spectral
  Codes}, Computer Physics Communications 177 (2007) 560--565.
\newblock \href {http://dx.doi.org/10.1016/j.cpc.2007.05.019}
  {\path{doi:10.1016/j.cpc.2007.05.019}}.

\bibitem{Bailey2005}
D.~H. Bailey, {High-Precision Floating-Point Arithmetic in Scientific
  Computing}, Computing in Science and Engineering 7~(3) (2005) 54--61.

\bibitem{Trojak2020}
W.~Trojak, A.~Scillitoe, R.~Watson,
  \href{https://arc.aiaa.org/doi/10.2514/6.2020-0566}{{Effect of Flux Function
  Order and Working Precision in Spectral Element Methods}}, in: AIAA Scitech
  2020 Forum, no. January, American Institute of Aeronautics and Astronautics,
  Orlando, FL, 2020, pp. 1--21.
\newblock \href {http://dx.doi.org/10.2514/6.2020-0566}
  {\path{doi:10.2514/6.2020-0566}}.
\newline\urlprefix\url{https://arc.aiaa.org/doi/10.2514/6.2020-0566}

\bibitem{Nvidia2017}
Nvidia, {Nvidai Tesla v100 Gpu Architecture; The Wolrd's Most Advanced Data
  Center Gpu}, Tech. Rep. August (2017).

\bibitem{IEEE754_2008}
D.~Zuras, M.~Cowlishaw, A.~Aiken, M.~Applegate, D.~Bailey, S.~Bass,
  D.~Bhandarkar, M.~Bhat, D.~Bindel, S.~Boldo, S.~Canon, S.~R. Carlough,
  M.~Cornea, J.~H. Crawford, J.~D. Darcy, D.~Das, S.~M. Daumas, B.~Davis,
  M.~Davis, D.~Delp, J.~Demmel, M.~A. Erle, H.~A.~H. Fahmy, J.~P. Fasano,
  R.~Fateman, E.~Feng, W.~E. Ferguson, A.~Fit-Florea, L.~Fournier, C.~Freitag,
  I.~Godard, R.~A. Golliver, D.~Gustafson, M.~Hack, J.~R. Harrison, J.~Hauser,
  Y.~Hida, C.~N. Hinds, G.~Hoare, D.~G. Hough, J.~Huck, J.~Hull, M.~Ingrassia,
  D.~V. James, R.~James, W.~Kahan, J.~Kapernick, R.~Karpinski, J.~Kidder,
  P.~Koev, R.-C. Li, Z.~A. Liu, R.~Mak, P.~Markstein, D.~Matula, G.~Melquiond,
  N.~Mori, R.~Morin, N.~Nedialkov, C.~Nelson, S.~Oberman, J.~Okada, I.~Ollmann,
  M.~Parks, T.~Pittman, E.~Postpischil, J.~Riedy, E.~M. Schwarz, D.~Scott,
  D.~Senzig, I.~Sharapov, J.~Shearer, M.~Siu, R.~Smith, C.~Stevens, P.~Tang,
  P.~J. Taylor, J.~W. Thomas, B.~Thompson, W.~Thrash, N.~Toda, S.~D. Trong,
  L.~Tsai, C.~Tsen, F.~Tydeman, L.~K. Wang, S.~Westbrook, S.~Winkler, A.~Wood,
  U.~Yalcinalp, F.~Zemke, P.~Zimmermann,
  \href{http://ieeexplore.ieee.org/xpl/freeabs{\_}all.jsp?arnumber=4610935{\%}5Cnhttp://ieeexplore.ieee.org/servlet/opac?punumber=4610933}{{IEEE
  Std 754™-2008 (Revision of IEEE Std 754-1985), IEEE Standard for
  Floating-Point Arithmetic}}, Tech. Rep. 754-2008 (2008).
\newblock \href {http://dx.doi.org/10.1109/IEEESTD.2008.4610935}
  {\path{doi:10.1109/IEEESTD.2008.4610935}}.
\newline\urlprefix\url{http://ieeexplore.ieee.org/xpl/freeabs{\_}all.jsp?arnumber=4610935{\%}5Cnhttp://ieeexplore.ieee.org/servlet/opac?punumber=4610933}

\bibitem{Delp1979}
E.~J. Delp, O.~R. Mitchell, {Image Compression Using Block Truncation Coding},
  IEEE Transactions on Coimmunication 27~(9) (1979) 1335--1342.
\newblock \href {http://arxiv.org/abs/arXiv:1011.1669v3}
  {\path{arXiv:arXiv:1011.1669v3}}, \href
  {http://dx.doi.org/10.1017/CBO9781107415324.004}
  {\path{doi:10.1017/CBO9781107415324.004}}.

\bibitem{Ning1992}
P.~Ning, L.~Hesselink, {Vector Quantization for Volume Rendering}, in: VVS '92
  Proceedings of the 1992 workshop on Volume visualization, 1992, pp. 69--74.

\bibitem{Campbell1986}
G.~Campbell, T.~A. DeFanti, J.~Frederiksen, S.~A. Joyce, L.~A. Leske, J.~A.
  Lindberg, D.~J. Sandin, {Two Bit/Pixel Full Color Encoding}, in: SIGGRAPH '86
  Proceedings of the 13th annual conference on Computer graphics and
  interactive techniques, Vol.~20, 1986, pp. 215--223.
\newblock \href {http://arxiv.org/abs/arXiv:1011.1669v3}
  {\path{arXiv:arXiv:1011.1669v3}}, \href
  {http://dx.doi.org/10.1017/CBO9781107415324.004}
  {\path{doi:10.1017/CBO9781107415324.004}}.

\bibitem{Linde1980}
Y.~Linde, A.~Buzo, R.~M. Gray, {An Algorithm for Vector Quantizer Design}, IEEE
  Transactions on Communications 28~(1) (1980) 84--95.
\newblock \href {http://dx.doi.org/10.1109/TCOM.1980.1094577}
  {\path{doi:10.1109/TCOM.1980.1094577}}.

\bibitem{Iourcha1999}
I.~K. Iourcha, K.~S. Nayak, Z.~Hong, {System and Method for Fixed-Rate
  Block-Based Image Compression with Inferred Pixel Values} (1999).

\bibitem{Schneider2003}
J.~Schneider, R.~Westermann, {Compression Domain Volume Rendering}, in:
  Proceedings of the IEEE Visualization Conference, 2003, pp. 293--300.
\newblock \href {http://dx.doi.org/10.1109/VISUAL.2003.1250385}
  {\path{doi:10.1109/VISUAL.2003.1250385}}.

\bibitem{Lindstrom2014a}
P.~Lindstrom, {Fixed-rate compressed floating-point arrays}, IEEE Transactions
  on Visualization and Computer Graphics 20~(12) (2014) 2674--2683.
\newblock \href {http://dx.doi.org/10.1109/TVCG.2014.2346458}
  {\path{doi:10.1109/TVCG.2014.2346458}}.

\bibitem{Alted2010}
F.~Alted, {Why modern CPUs are starving and what can be done about it},
  Computing in Science and Engineering 12~(2) (2010) 68--71.
\newblock \href {http://dx.doi.org/10.1109/MCSE.2010.51}
  {\path{doi:10.1109/MCSE.2010.51}}.

\bibitem{Ziv1977}
J.~Ziv, A.~Lempel, {A Universal Algorithm for Sequential Data Compression},
  IEEE Transactions on Information Theory 23~(3) (1977) 337--343.
\newblock \href {http://dx.doi.org/10.1109/TIT.1977.1055714}
  {\path{doi:10.1109/TIT.1977.1055714}}.

\bibitem{Ziv1978}
J.~Ziv, A.~Lempel, {Compression of lndiwdual Sequences}, IEEE Transactions on
  Information Theory 24~(5) (1978) 530--536.

\bibitem{Cigolle2014}
Z.~H. Cigolle, M.~Mara, S.~Donow, M.~Mcguire, D.~Evangelakos, Q.~Meyer, {A
  Survey of Efficient Representations for Independent Unit Vectors}, Journal of
  Computer Graphics Techniques 3~(2) (2014) 1--30.

\bibitem{Meyer2010}
Q.~Meyer, J.~S{\"{u}}{\ss}muth, G.~Su{\ss}ner, M.~Stamminger, G.~Greiner, {On
  floating-point normal vectors}, Computer Graphics Forum 29~(4) (2010)
  1405--1409.
\newblock \href {http://dx.doi.org/10.1111/j.1467-8659.2010.01737.x}
  {\path{doi:10.1111/j.1467-8659.2010.01737.x}}.

\bibitem{Smith2012a}
J.~Smith, G.~Petrova, S.~Schaefer,
  \href{http://dx.doi.org/10.1016/j.cag.2012.03.017}{{Encoding normal vectors
  using optimized spherical coordinates}}, Computers and Graphics 36~(5) (2012)
  360--365.
\newblock \href {http://dx.doi.org/10.1016/j.cag.2012.03.017}
  {\path{doi:10.1016/j.cag.2012.03.017}}.
\newline\urlprefix\url{http://dx.doi.org/10.1016/j.cag.2012.03.017}

\bibitem{Bertsimas1993}
D.~Bertsimas, J.~Sitsiklis,
  \href{http://www.mit.edu/{~}dbertsim/papers/Optimization/Simulated
  annealing.pdf}{{Simulated Annealing}} (1993).
\newline\urlprefix\url{http://www.mit.edu/{~}dbertsim/papers/Optimization/Simulated
  annealing.pdf}

\bibitem{Huynh2007}
H.~T. Huynh, \href{http://arc.aiaa.org/doi/pdf/10.2514/6.2007-4079}{{A Flux
  Reconstruction Approach to High-Order Schemes Including Discontinuous
  Galerkin Methods}}, in: 18th AIAA Computational Fluid Dynamics Conference,
  Vol. 2007-4079, 2007, pp. 1--42.
\newblock \href {http://dx.doi.org/10.2514/6.2007-4079}
  {\path{doi:10.2514/6.2007-4079}}.
\newline\urlprefix\url{http://arc.aiaa.org/doi/pdf/10.2514/6.2007-4079}

\bibitem{Huynh2009}
H.~T. Huynh, \href{http://arc.aiaa.org/doi/pdf/10.2514/6.2007-4079
  http://dx.doi.org/10.2514/6.2007-4079}{{A Flux Reconstruction Approach to
  High-Order Schemes Including Discontinuous Galerkin for Diffusion}}, in: 47th
  AIAA Aerospace Science Meeting, no. January in Fluid Dynamics and Co-located
  Conferences, American Institute of Aeronautics and Astronautics, 2009, pp.
  1--34.
\newblock \href {http://dx.doi.org/doi:10.2514/6.2007-4079}
  {\path{doi:doi:10.2514/6.2007-4079}}.
\newline\urlprefix\url{http://arc.aiaa.org/doi/pdf/10.2514/6.2007-4079
  http://dx.doi.org/10.2514/6.2007-4079}

\bibitem{Vincent2010}
P.~E. Vincent, P.~Castonguay, A.~Jameson,
  \href{http://link.springer.com/10.1007/s10915-010-9420-z}{{A New Class of
  High-Order Energy Stable Flux Reconstruction Schemes}}, Journal of Scientific
  Computing 47~(1) (2010) 50--72.
\newblock \href {http://dx.doi.org/10.1007/s10915-010-9420-z}
  {\path{doi:10.1007/s10915-010-9420-z}}.
\newline\urlprefix\url{http://link.springer.com/10.1007/s10915-010-9420-z}

\bibitem{Rusanov1961}
V.~Rusanov, {The Calculation of the Interaction of Non-Stationary Shock Waves
  with Barriers}, Zh. Vychisl. Mat. Mat. Fiz. 1~(2) (1961) 267--279.
\newblock \href {http://arxiv.org/abs/arXiv:1011.1669v3}
  {\path{arXiv:arXiv:1011.1669v3}}, \href
  {http://dx.doi.org/10.18287/0134-2452-2015-39-4-453-458.}
  {\path{doi:10.18287/0134-2452-2015-39-4-453-458.}}

\bibitem{Davis1988}
S.~Davis, {Simplified Second-order Godunov-type Methods}, SIAM Journal on
  Scientific and Statistical Computing 9~(3) (1988) 445--473.

\bibitem{Bassi1997a}
F.~Bassi, S.~Rebay,
  \href{http://linkinghub.elsevier.com/retrieve/pii/S0021999196955722}{{A
  High-Order Accurate Discontinuous Finite Element Method for the Numerical
  Solution of the Compressible Navier–Stokes Equations}}, Journal of
  Computational Physics 131~(2) (1997) 267--279.
\newblock \href {http://dx.doi.org/10.1006/jcph.1996.5572}
  {\path{doi:10.1006/jcph.1996.5572}}.
\newline\urlprefix\url{http://linkinghub.elsevier.com/retrieve/pii/S0021999196955722}

\bibitem{Kennedy2000}
C.~A. Kennedy, M.~H. Carpenter, R.~M. Lewis, {Low-Storage, Explicit Runge-Kutta
  Schemes for the Compressible Navier-Stokes Equations}, Applied Numerical
  Mathematics 35~(1) (2000) 177--219.

\bibitem{Romero2020}
J.~Romero, J.~Crabill, J.~Watkins, F.~Witherden, A.~Jameson,
  \href{https://doi.org/10.1016/j.cpc.2020.107169}{{ZEFR}: A {GPU}-accelerated
  high-order solver for compressible viscous flows using the flux
  reconstruction method}, Computer Physics Communications 250 (2020) 107169.
\newblock \href {http://dx.doi.org/10.1016/j.cpc.2020.107169}
  {\path{doi:10.1016/j.cpc.2020.107169}}.
\newline\urlprefix\url{https://doi.org/10.1016/j.cpc.2020.107169}

\bibitem{Shu1997}
C.-w. Shu, {Essentially Non-Oscillatory and Weighted Essentially
  Non-Oscillatory Schemes for Hyperbolic Conservation Laws Operated by
  Universities Space Research Association}, ICASE Report~(97-65) (1997) 1--78.
\newblock \href {http://dx.doi.org/10.1007/BFb0096355}
  {\path{doi:10.1007/BFb0096355}}.

\bibitem{Taylor1937}
G.~I. Taylor, A.~E. Green,
  \href{http://rspa.royalsocietypublishing.org/cgi/doi/10.1098/rspa.1937.0036}{{Mechanism
  of the Production of Small Eddies from Large Ones}}, Proceedings of the Royal
  Society A: Mathematical, Physical and Engineering Sciences 158~(895) (1937)
  499--521.
\newblock \href {http://arxiv.org/abs/arXiv:1205.0516v2}
  {\path{arXiv:arXiv:1205.0516v2}}, \href
  {http://dx.doi.org/10.1098/rspa.1937.0036}
  {\path{doi:10.1098/rspa.1937.0036}}.
\newline\urlprefix\url{http://rspa.royalsocietypublishing.org/cgi/doi/10.1098/rspa.1937.0036}

\bibitem{Brachet1983}
M.~E. Brachet, S.~A. Orszag, B.~G. Nickel, R.~H. Morf, U.~Frisch, {Small-Scale
  Structure of the Taylor-Green Vortex}, Journal of Fluid Mechanics 130 (1983)
  411--452.

\bibitem{DeBonis2013}
J.~R. DeBonis, \href{http://arc.aiaa.org/doi/10.2514/6.2013-382}{{Solutions of
  the Taylor-Green Vortex Problem Using High-Resolution Explicit Finite
  Difference Methods}}, 51st AIAA Aerospace Sciences Meeting including the New
  Horizons Forum and Aerospace Exposition~(February 2013).
\newblock \href {http://dx.doi.org/10.2514/6.2013-382}
  {\path{doi:10.2514/6.2013-382}}.
\newline\urlprefix\url{http://arc.aiaa.org/doi/10.2514/6.2013-382}

\bibitem{Trojak2018d}
W.~Trojak, R.~Watson, A.~Scillitoe, P.~G. Tucker,
  \href{https://arxiv.org/pdf/1809.05189.pdf}{{Effect of Mesh Quality on Flux
  Reconstruction in Multi-Dimensions}}, Tech. rep. (2018).
\newblock \href {http://arxiv.org/abs/arXiv:1809.05189v1}
  {\path{arXiv:arXiv:1809.05189v1}}.
\newline\urlprefix\url{https://arxiv.org/pdf/1809.05189.pdf}

\end{thebibliography}


\clearpage
\begin{appendices}

\end{appendices}


\end{document}